\documentclass[aps,prl,reprint,nobibnotes,
longbibliography,superscriptaddress]{revtex4-1}
\usepackage{t1enc}
\usepackage[utf8]{inputenc}
\usepackage{amsmath}
\usepackage{graphicx}
\usepackage{enumerate}
\usepackage{natbib}
\usepackage{multibib}
\usepackage{float}
\usepackage{color}
\newcites{M}{references_ABS_nourl}
\newcites{S}{references_ABS_nourl_suppl}

\renewcommand{\Re}{\textrm{Re}}

\newcommand{\bra}[1]{\langle #1|}
\newcommand{\ket}[1]{|#1\rangle}
\newcommand{\braket}[2]{\langle #1|#2\rangle}

\newcommand{\abs}[1]{\left\vert#1\right\vert}

\newcommand{\pd}{\partial}

\DeclareMathOperator{\de}{d\!}

\begin{document}

\title{Microwave spectroscopy of spinful Andreev bound states in ballistic
semiconductor Josephson junctions}

\author{David J.~van Woerkom}

\author{Alex Proutski}

\affiliation{QuTech, Delft University of
Technology, 2600 GA Delft, The Netherlands}

\affiliation{Kavli Institute of Nanoscience, Delft University of
Technology, 2600 GA Delft, The Netherlands}

\author{Bernard van Heck}

\affiliation{Department of Physics, Yale University, New Haven, CT 06520, USA}

\author{Dani{\"e}l Bouman}

\affiliation{QuTech, Delft University of
Technology, 2600 GA Delft, The Netherlands}

\affiliation{Kavli Institute of Nanoscience, Delft University of
Technology, 2600 GA Delft, The Netherlands}

\author{Jukka I. V{\"a}yrynen}

\author{Leonid I. Glazman}

\affiliation{Department of Physics, Yale University, New Haven, CT 06520, USA}

\author{Peter Krogstrup}

\author{Jesper Nyg{\aa}rd}

\affiliation{Center for Quantum Devices and Station Q Copenhagen, Niels Bohr Institute,
University of Copenhagen, Universitetsparken 5, 2100 Copenhagen, Denmark}

\author{Leo P. Kouwenhoven}

\author{Attila Geresdi}

\affiliation{QuTech, Delft University of
Technology, 2600 GA Delft, The Netherlands}

\affiliation{Kavli Institute of Nanoscience, Delft University of
Technology, 2600 GA Delft, The Netherlands}

\maketitle

\textbf{The superconducting proximity effect in semiconductor nanowires
 has recently enabled the study of new superconducting
architectures, such as gate-tunable superconducting qubits and multiterminal
Josephson junctions. As opposed to their metallic counterparts, the electron
density in semiconductor nanosystems is tunable by external electrostatic gates
providing a highly scalable and \emph{in-situ} variation of the device
properties. In addition, semiconductors with large $g$-factor and spin-orbit
coupling have been shown to give rise to exotic phenomena in superconductivity, such as
$\varphi_0$ Josephson junctions and the emergence of Majorana bound states.
Here, we report microwave spectroscopy measurements that directly reveal the
presence of Andreev bound states (ABS) in ballistic semiconductor channels. We
show that the measured ABS spectra are the result of transport channels with
gate-tunable, high transmission probabilities up to $0.9$, which is required for
gate-tunable Andreev qubits and beneficial for braiding schemes of Majorana
states. For the first time, we detect excitations of a spin-split pair of ABS 
and observe symmetry-broken ABS, a direct consequence of the spin-orbit coupling
in the semiconductor.}

\begin{figure}[!ht]
\centering
\includegraphics[width=0.5\textwidth]{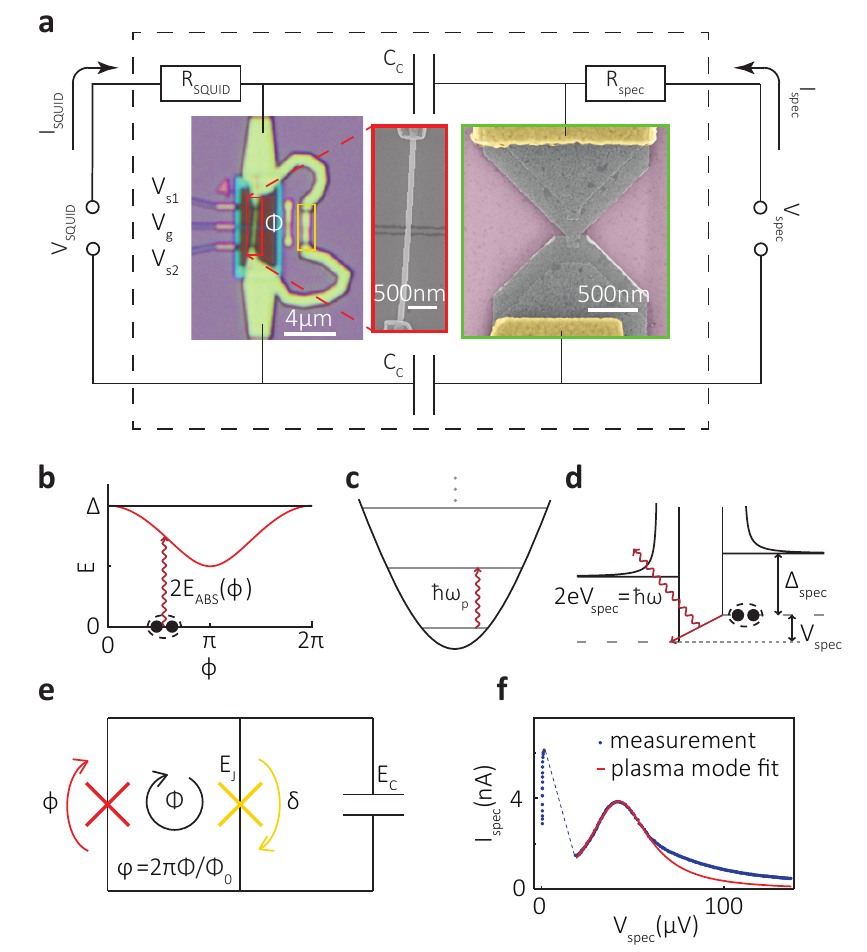}
\caption{\textbf{Device schematics and working principle.} (a) Equivalent
circuit diagram: Bright field optical image of the hybrid SQUID with one InAs
semiconductor nanowire weak link (scanning electron micrograph, in the red box)
and an Al/AlO$_x$/Al tunnel junction (enclosed by the yellow box). The SQUID is
capacitively coupled to the spectrometer Al/AlO$_x$/Al Josephson junction
(scanning electron micrograph, in the green box) via $C_c$. The transmission of
the semiconductor channel is tuned by the gate voltage, $V_g$. Additional gates near
the electrodes are kept at a constant voltage $V_{s1,2}$. Circuit elements
within the dashed box are located on-chip, thermally anchored to
$12\,$mK. (b) and (c) excitations of the hybrid SQUID: the
Andreev bound state at $\hbar\omega = 2E_\textrm{ABS}$ (b) and the plasma
oscillations at $\hbar\omega = \hbar\omega_p$ (c) are excited by a photon energy
$\hbar\omega = 2eV_\textrm{spec}$ set by the DC voltage bias of the
spectrometer (d) with a superconducting gap $\Delta_\textrm{spec}$. (e)
Schematic circuit diagram of the hybrid SQUID.
The total phase $\varphi=\phi + \delta$ is determined by the applied flux $\Phi$. (f) The
measured $I(V)$ trace of the spectrometer junction with the nanowire in full
depletion, i.e.~in the absence of ABS excitations. The red solid line shows the
fit to the circuit model of a single resonance centered at $\hbar \omega_p$, see
text. Images and data were all taken on device 1.}
\end{figure}

The linear conductance $G=\frac{2e^2}{h} \sum T_i$ of a nanostructure between
two bulk leads \citeM{LANDAUER198191} depends on the individual channel
transmission probabilities, $T_i$. Embedding the same structure between two
superconducting banks with a superconducting gap of $\Delta$ gives rise to
Andreev bound states (ABS) \citeM{kulik1969}. If the junction length is much
smaller than the superconducting coherence length, $\xi$, i.e.~in the short
junction limit, then the ABS levels depend on the phase difference $\phi$
between the leads according to \citeM{PhysRevLett.67.3836}:
\begin{equation}
E_\textrm{ABS,i}(\phi) = \pm \Delta \sqrt{1-T_i\sin^2{\frac{\phi}{2}}}\,.
\end{equation}
These subgap states with $\left | E_\textrm{ABS} \right | \le \Delta$ are
localized in the vicinity of the nanostructure and extend into the banks
over a length scale determined by $\xi$.
Note that Eq.~(1) is only valid in the absence of magnetic field, when each
energy level is doubly degenerate.

\begin{figure*}[!ht]
\centering
\includegraphics[width=1.0\textwidth]{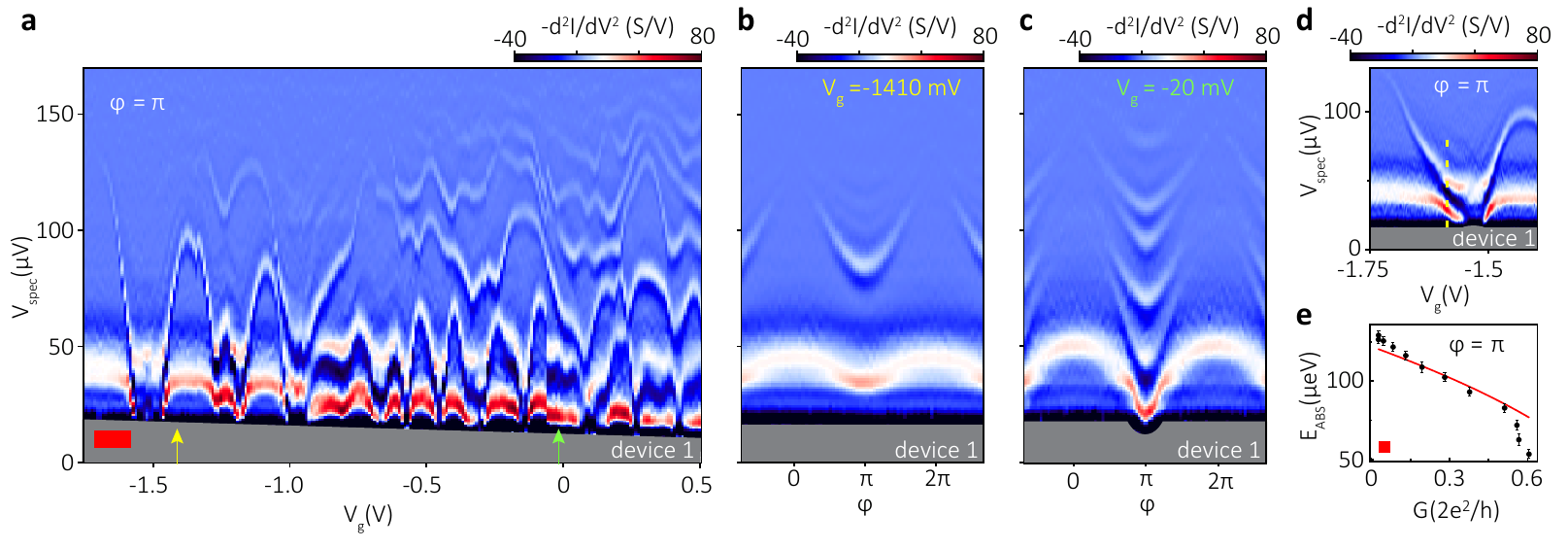}
\caption{\textbf{Gate dependence of Andreev bound states.}
(a) $-d^2I/dV^2$ of the spectrometer junction as a function of $V_g$ at
$\varphi=\pi$, where $E_\textrm{ABS,i}=\Delta\sqrt{1-T_i}$ in the short junction
limit. Panels (b) and (c): $-d^2I/dV^2$ of the spectrometer junction as a
function of $\varphi=2\pi \Phi/\Phi_0$ for one channel (b) and several channels
(c). The qualitative agreement of the line shapes with Eq.~(1) confirms the
short junction behaviour. Arrows in panel (a) indicate $V_g$ for these measurements.
Weakly visible vertically shifted replicas of the ABS lines indicate higher
order transitions, see text.
(d) Strong hybridization between the ABS excitation and the plasma mode with a
level repulsion of $\varepsilon=22\,\mu$eV at the yellow dashed line. (e)
$E_\textrm{ABS}(\varphi=\pi)$ as a function of the DC linear conductance $G$ of
the nanowire weak link in the gate span denoted by the red bar in panel (a). The
error bars correspond to the linewidth of the measured signal.
The solid red line shows the prediction of the single channel model with
$\Delta=122\,\mu$eV$\pm3\,\mu$eV, see text. All data was taken on device 1.
Grey regions denote lack of data due to bias instability of the circuit.}
\end{figure*}

Direct microwave spectroscopy has recently demonstrated the occupation of the
ABS by exciting a Cooper pair in atomic junctions \citeM{bretheau_nature}.
Unlike quasiparticle tunneling spectroscopy, which has also been used to detect
ABS \citeM{pillet_CNT_ABS, PhysRevLett.110.217005}, resonant excitation by
microwaves is a charge parity-conserving process \citeM{PhysRevB.87.174521}. This
property enables coherent control of ABS which is required for novel qubit architectures
\citeM{Janvier1199} and makes microwave spectroscopy a promising tool to detect
Majorana bound states \citeM{PhysRevB.92.134508} in proximitized semiconductor systems
\citeM{PhysRevLett.105.077001, PhysRevLett.105.177002, Mourik1003}.

We investigate ABS excitations in Josephson junctions that consist of indium
arsenide (InAs) nanowires covered by epitaxial aluminium (Al) shells
\citeM{CPH_InAs_nmat}. The junction, where the superconducting shell is removed,
is $100\,$nm (device 1, see the red box in Fig.~1a) and $40\,$nm long (device
2), respectively. The nanowire is then embedded in a hybrid superconducting
quantum interference device (SQUID) whose second arm is a conventional
Al/AlO$_x$/Al tunnel junction (in yellow box), enabling the control of the phase
drop $\phi$ by means of the applied magnetic flux $\Phi$ through the SQUID loop.
In the limit of a negligible loop inductance and an asymmetric SQUID, where the
Josephson coupling of the nanowire is much smaller than that of the tunnel
junction, the applied phase $\varphi$ mostly drops over the nanowire link:
$\phi\approx \varphi=2\pi \Phi/\Phi_0$, where $\Phi_0=h/2e$ is the
superconducting flux quantum. We measure the microwave response
\citeM{bretheau_nature, PhysRevB.87.174521} of the nanowire junction utilizing
the circuit depicted in Fig.~1a, where a second Al/AlO$_x$/Al tunnel junction (in green box)
is capacitively coupled to the hybrid SQUID and acts as a spectrometer. Further
details on the fabrication process are given in the Supplementary Material.

In this circuit, inelastic Cooper-pair tunneling (ICPT, Fig.~1d) of the
spectrometer junction is enabled by the dissipative environment and
results in a DC current, $I_\textrm{spec}$ \citeM{PhysRevLett.73.3455}:
\begin{equation}
I_\textrm{spec} = \frac{I_\textrm{c,spec}^2\Re[Z(\omega)]}{2V_\textrm{spec}}.
\end{equation}
Here $I_\textrm{c,spec}$ is the critical current of the spectrometer junction,
$V_\textrm{spec}$ is the applied voltage bias, and $Z(\omega)$ is the circuit
impedance at a frequency $\omega=2eV_\textrm{spec}/\hbar$. Since $Z(\omega)$
peaks at the resonant frequencies of the hybrid SQUID
\citeM{PhysRevLett.73.3455, bretheau_nature}, so does the DC current
$I_\textrm{spec}$, allowing us to measure the ABS excitation energies of the
nanowire junction (Fig.~1b), as well as the plasma frequency of the SQUID
(Fig.~1c).

First we characterize the contribution of the plasma mode with the nanowire
junction gated to full depletion, i.e.~$G=0$. We show the
$I(V)$ curve of the spectrometer junction of device 1 in Fig.~1f, where we find a
single peak centered at $\hbar\omega_p/2=eV_\textrm{spec} = 46\,\mu$eV and a
quality factor $Q\approx 1$. In the limit of $E_C \ll E_J$,
$\hbar\omega_p=\sqrt{2E_C E_J}$, where $E_C$ is the charging energy of the
circuit and $E_J$ is the Josephson coupling of the tunnel junction (Fig.~1e).
Estimating $E_J=165\,\mu$eV from the normal state resistance
\citeM{PhysRevLett.10.486}, this measurement allows us to determine
$E_C=25.4\,\mu$eV (see the Supplementary Material). The choice of a low quality
factor in combination with a characteristic impedance $Z_0 = 551\,\Omega \ll
R_q=h/4e^2$ ensures the suppression of higher order transitions and parasitic
resonances.

Next, we investigate the spectrometer response as a function of the gate voltage
$V_g$ applied to the nanowire. Note that the spectrometer response to the ABS
transitions is superimposed on the plasma resonance peak. In order to achieve a
better visibility of the ABS lines, we display $-d^2
I_\textrm{spec}/dV_\textrm{spec}^2(V_\textrm{spec})$ rather than
$I_\textrm{spec}(V_\textrm{spec})$ (see Supplementary Material for comparison).
In the presence of ABS, the spectrum
exhibits peaks at frequencies where $\hbar \omega = 2E_{\textrm{ABS},i}$
\citeM{PhysRevB.87.174521}. In Fig.~2a, we monitor the appearance of these peaks
for an applied phase $\varphi=\pi$, where the ABS energy of Eq.~(1) is
$E_{\textrm{ABS},i}(\pi)=\Delta \sqrt{1-T_i}$. Notably, for $V_g$ values close
to full depletion (see red bar in Fig.~2a), we see a gradual decrease of
$E_{\textrm{ABS}}(\pi)$ with increasing $V_g$ (black circles in Fig.~2e). In
this regime, we find a good correspondence with Eq.~(1), assuming single channel transport,
$G=\frac{2e^2}{h}T$ (red solid line in Fig.~2e, see the Supplementary material
on the details of the measurement of $G$).
However, the observed $\Delta=122\,\mu$eV is smaller than the
$\Delta_\textrm{Al}\approx 200\,\mu$eV of the thin film Al contacts, in
agreement with the presence of induced superconductivity in the nanowire
\citeM{CPH_InAs_nnano}. Increasing $V_g$ further, we observe a sequential
appearance of peaks, which we attribute to the opening of multiple transport
channels in the weak link and the consequent formation of multiple ABS
\citeM{PhysRevLett.67.3836} as the Fermi level, $E_F$ increases. We also find a
strong variation of $E_\textrm{ABS}$ with $V_g$ similarly to earlier experiments \citeM{Doh_2006,
PhysRevLett.115.127001, PhysRevLett.115.127002}. We attribute this observation
to mesoscopic fluctuations in the presence of weak disorder
\citeM{PhysRevLett.67.3836}, such that the mean free path of the charge carriers
is comparable to the channel length.

Now we turn to the flux dependence of the observed spectrum, shown in Fig.~2b
and 2c for two distinct gate configurations. We find a qualitative agreement
with Eq.~(1) with one transport channel in Fig.~2b and several channels in
Fig.~2c confirming that our device is in the short junction limit. In addition,
we observe the plasma mode at $eV_\textrm{spec} < 50\,\mu$eV. We also find
that the plasma mode $\hbar\omega_p$ oscillates with $\varphi$ when the nanowire
is gated to host open transport channels. This is expected due to the Josephson
coupling of the nanowire becoming comparable to $E_J$, which also causes a
finite phase drop, $\delta$, over the tunnel junction (see Supplementary
Material). We also note the presence of additional, weakly visible lines in the
spectrum which could be attributed to higher order processes
\citeM{bretheau_nature}. However, we did not identify the nature of these
excitations, and we focus on the main transitions throughout the current
work.

In addition, we observe the occurrence of avoided crossings between the Andreev
and plasma modes, as shown in Fig.~2d at $\varphi=\pi$.
These avoided crossings require $\hbar\omega_p \approx 2 \Delta\sqrt{1-T}$,
which translates to a high transmission probability $T\approx 0.8-0.9$, and
demonstrates the hybridization between the ABS excitation and the plasma mode.
The coupling between these two degrees of freedom has previously been derived
\citeM{PhysRevB.87.174521, PhysRevB.90.134506} (see Supplementary Material),
leading to a perturbative estimate for the energy splitting $\varepsilon \approx
\Delta T \left (E_C/2E_J \right )^{1/4} \approx 40-70\,\mu$eV, similar to
the observed value of $22\,\mu$eV. The discrepancy is fully resolved in the
numerical analysis of the circuit model developed below.

\begin{figure}
\centering
\includegraphics[width=0.5\textwidth]{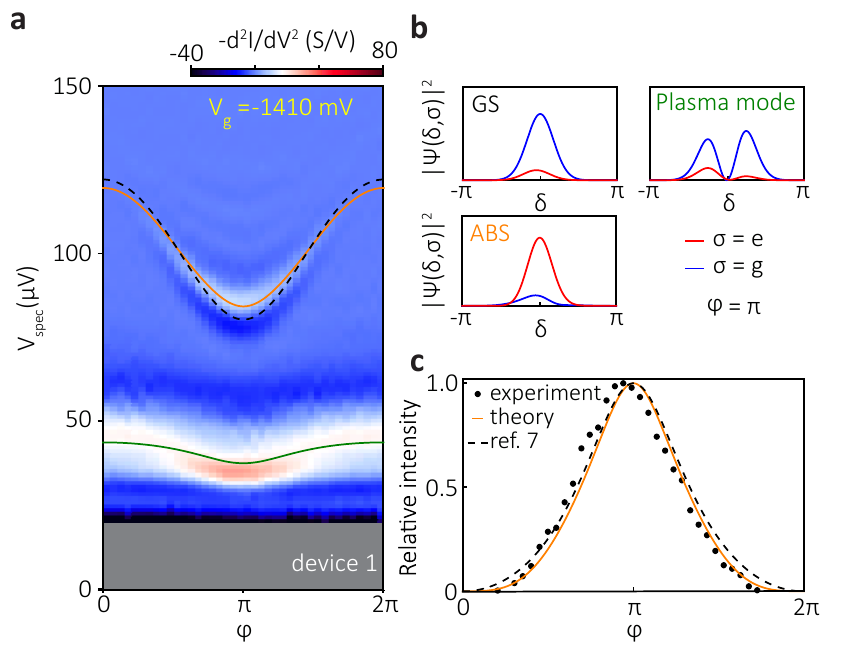}
\caption{\textbf{Theoretical description of the transitions.}
(a) Solid lines denote the transitions identified by the model described in the
text, with $\Delta$ and $T$ being free parameters.
The experimental dataset is the same as the one shown in Fig.~2b. The dashed
line shows Eq.~(1) for the fitted
$\Delta=122\,\mu$eV and $T=0.57$. (b) The probability density
$\left |\Psi(\delta, \sigma)\right |^2$ in the ground state of the hybrid SQUID
(GS), and in the two excited states depicted in panel (a), respectively. The weight in the
ABS ground state ($\sigma=g$) and in the ABS excited state ($\sigma=e$)
distinguishes between the plasma mode and the ABS. (c) The
measured relative intensity of the ABS transition (black circles) compared to
the theoretical expectation based on Eq.~(3) (orange solid line) and from
\protect\citeM{PhysRevB.87.174521} (black dashed line) with no additional
fitting parameters.}
\end{figure}

We provide a unified description of
the energy spectrum of the circuit as a whole, and consider the following
Hamiltonian for the hybrid SQUID (Fig.~1e) \citeM{PhysRevB.90.134506}:
\begin{equation}
\hat{H} = E_C \hat{N}^2 + E_J(1-\cos\hat{\delta}) +
\hat{H}_\textrm{ABS}(\varphi-\hat{\delta})\,.
\end{equation}
Here $\hat{\delta}$ is the operator of the phase difference across the tunnel
junction, conjugate to the charge operator $\hat{N}$, $[\hat{\delta}, \hat{N}]=i$.
The first two terms in Eq.~(3) represent the charging energy of the circuit and
the Josephson energy of the tunnel junction (Fig.~1e). The last term describes
the quantum dynamics of a single-channel short weak link
\citeM{PhysRevLett.90.087003, PhysRevB.71.214505}, which depends on $\Delta$
and $T$. For the analytic form of $\hat{H}_\textrm{ABS}$, see the Supplementary Material. To
fully account for the coupling between the ABS excitation and the quantum
dynamics of the phase across the SQUID, we numerically solve the eigenvalue
problem $\hat{H}\,\Psi = E\,\Psi$ and determine the transition frequencies
$\hbar\omega = E-E_\textrm{GS}$ with $E_\textrm{GS}$ being the ground state energy.

This procedure allows us to fit the experimental data, and we find a good
quantitative agreement as shown in Fig.~3a for a dataset taken at
$V_g=-1410\,$mV with the fit parameters $\Delta=122\,\mu$eV and $T=0.57$. The
previously identified circuit parameters $E_J$ and $E_C$ are kept fixed during
the fit. We note that the observed ABS transition (orange solid line) only
slightly deviates from Eq.~(1) (black dashed line). The modulation of the plasma
frequency (green solid line) is then defined by the model Hamiltonian with no
additional fit parameters. We further confirm the nature of the plasma and ABS
excitations by evaluating the probability density $\left |\Psi(\delta, \sigma)
\right|^2$ of the eigenfunctions of Eq.~(3) at $\varphi=\pi$ (Fig.~3b). In the
ground state of $\hat{H}$ (GS) and in the state corresponding to the plasma
excitation (green line in Fig.~3a), the probability density is much higher in
the ground state of the weak link ($\sigma=g$, blue line) than in the excited
state ($\sigma=e$, red line). In contrast, the next observed transition (orange
line in Fig.~3a) gives rise to a higher contribution from $\sigma=e$ confirming
our interpretation of the experimental data in terms of ABS excitations.
Furthermore, the model can also describe measurement data with $T$ close to
1, where it accurately accounts for the avoided crossings between the ABS and
plasma spectral lines (see the Supplementary Material for a dataset with
$T=0.9$).

In Fig.~3c we show the visibility of the ABS transition as a function of
the applied phase $\varphi$, which is proportional to the absorption
rate of the weak link, predicted to be $\propto T^2 (1-T)
\sin^4(\varphi/2)\,\times \Delta^2/E_\textrm{ABS}^2(\varphi)$
\citeM{PhysRevB.87.174521}. We note that in the experimental data the maximum of
the intensity is slightly shifted from its expected position at $\varphi=\pi$.
This minor deviation may stem from the uncertainty of the flux calibration.
Nevertheless, using $T=0.57$, obtained from the fit in Fig.~3a, we find a good
agreement with no adjustable parameters (black dashed line). A similarly good
correspondence is also found with the full numerical model (orange line) based
on Eq.~(3).

\begin{figure}
\centering
\includegraphics[width=0.5\textwidth]{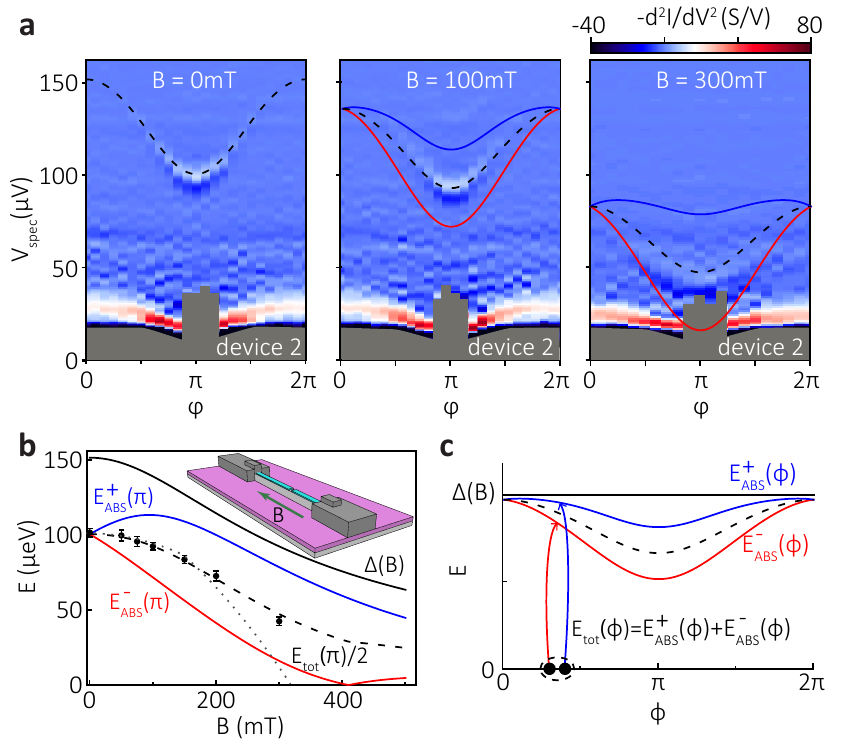}
\caption{\textbf{Spectroscopy of spin-split Andreev bound states in a Rashba
nanowire.} Panel (a) shows the flux dependence of the Andreev bound states at
$B=0$, $100$ and $300\,$mT, respectively, applied parallel to the nanowire. The
zero-field fit yields to $T=0.56$ and $\Delta=152\,\mu$eV. Dash lines depict the
fit of $E_\textrm{tot}(\phi)=E_\textrm{ABS}^+(\phi)+E_\textrm{ABS}^-(\phi)$ to
the model described in the text. (b) Black circles show the measured
$E_\textrm{tot}(\pi)$ as a function of $B$. The
error bars correspond to the linewidth of the measured signal.
The dashed line depicts the fit to the theory with $g=14.7\pm0.6$ and
$\sqrt{E_\textrm{SO}E_F}/\Delta = 0.32 \pm 0.02$, see text. The Zeeman-split ABS
levels $E_\textrm{ABS}^\pm(\pi)$ and the proximity-induced gap $\Delta(B)$
obtained from the model are shown as visual guides. The dotted line depicts the
expected behavior of $E_\textrm{tot}(B)$ in the presence of a strong orbital
magnetic field with $B_*=400\,$mT and weak spin-orbit coupling, see text. (c)
$E_\textrm{ABS}^\pm(\phi)$ computed at $B=100\,$mT are shown as blue and red
solid lines, together with the calculated transition energy
$E_\textrm{tot}(\phi)$ (black dashed line). The experimental data was taken on
device 2 at $V_g=140\,$mV. Grey regions denote lack of data due to bias
instability of the circuit.}
\end{figure}

We now discuss the evolution of the ABS as a function of an in-plane magnetic
field $B$ aligned parallel to the nanowire axis, which is perpendicular to the
internal Rashba spin-orbit field (see the inset in Fig.~4b for measurement
geometry). The applied field lifts the Kramers degeneracy of the energy
spectrum, splitting each Andreev doublet into a pair $E^\pm_\textrm{ABS}(\phi)$.
For small $B$, the splitting $E^+_\textrm{ABS}(\phi)-E^-_\textrm{ABS}(\phi)$ is
linear in $B$, due to the Zeeman effect. However, the spin-split single particle
levels are not accessible by microwave spectroscopy, which can only induce
transitions to a final state with two excited quasiparticles. Thus we can only
measure $E_\textrm{tot}(\phi)=E^+_\textrm{ABS}(\phi)+E^-_\textrm{ABS}(\phi)$ and
expect no split of the measured spectral lines. The experimental
data (Fig.~4a) shows that $E_\textrm{tot}$ decreases with $B$, while the
lineshape remains qualitatively intact.

In order to explain the field dependence of $E_\textrm{tot}$, we study the
behaviour of ABS in a simple model consisting of a short Josephson junction in a
one-dimensional quantum wire with proximity-induced superconductivity, Rashba
spin-orbit and an applied Zeeman field parallel to the wire
\citeM{PhysRevLett.105.077001,PhysRevLett.105.177002,PhysRevB.86.134522}. Within
this model, we are able to find $E^+_\textrm{ABS}$ and $E^-_\textrm{ABS}$, and reproduce the
observed quadratic decrease of the measured $E_\textrm{tot}(\pi)$ (black circles
in Fig.~4b). Initially, as $B$ is increased, the proximity-induced gap $\Delta(B)$
is suppressed (black solid line), while the energy $E_\textrm{ABS}^+(\pi)$ (blue
solid line) increases due to the Zeeman split of the ABS. However, a crossing of
the discrete ABS level with the continuum is avoided due to the presence of
spin-orbit coupling, which prevents level crossings in the energy spectrum by
breaking spin-rotation symmetry. The repulsion between the ABS level and the
continuum causes a downward bending of $E^+_\textrm{ABS}(\pi)$, in turn causing
a decrease in $E_\textrm{tot}(\pi)$ (black dashed line).

We perform the calculations in the limit where the Fermi level
$E_F$ in the wire is well above the Zeeman energy $E_Z = \frac{1}{2}g\mu_B B$
and the spin-orbit energy $E_\textrm{SO}=m\alpha^2/2\hbar^2$ with $m$ the
effective mass and $\alpha$ the Rashba spin-orbit coupling constant.
In this case and in the short junction limit, the ratio
$E_\textrm{tot}(\pi)/\Delta$ is a function of just two dimensionless parameters:
$E_Z/\Delta$ and $\sqrt{E_\textrm{SO}E_F}/\Delta$.
First we extract $\Delta=152\,\mu$eV and $T=0.56$ at $B=0$ (leftmost panel in
Fig.~4a). Then we perform a global fit on $E_\textrm{tot}(\phi)$ at all $B$
values and obtain a quantitative agreement with the theory for $g = 14.7\pm
0.6$, which is in line with expected $g$-factor values in InAs nanowires
\citeM{das_2012, higginbotham_natphys_2015,albrecht_nature_2016} and
$\sqrt{E_\textrm{SO}E_F}/\Delta = 0.32 \pm 0.02$.
This model is consistent assuming $E_F > E_Z \approx 100\,\mu$eV at $300\,$mT.
Thus we attain an upper bound $E_\textrm{SO}\lesssim 24\,\mu$eV, equivalent to a
Rashba parameter $\alpha\lesssim 0.12\,$eV{\AA} in correspondence with earlier
measurements on the same nanowires \citeM{albrecht_nature_2016}. However,
assuming the opposite limit, $E_F\approx 0$, the theory is not in agreement with
the experimental data (see the Supplementary Material).

The theoretical energy spectrum shown in Fig.~4b predicts a ground state
fermion-parity switch of the junction at a field $B_\textrm{sw}\approx
400\,$mT, at which the lowest ABS level $E^-_\textrm{tot}(\pi)=0$ (red line in
Fig.~4b). This parity switch inhibts the resonant excitation of the
Zeeman-split ABS levels \citeM{PhysRevB.77.184506} thus preventing microwave
spectroscopy measurements for $B>B_\textrm{sw}$. This prediction is in agreement
with the vanishing visibility of the ABS line at $B\approx B_\textrm{sw}$ in the
experiment.

In addition to the interplay of spin-orbit and Zeeman couplings, the orbital
effect of the magnetic field \citeM{PhysRevB.93.235434} is a second possible
cause for the decrease of the ABS transition energy.
Orbital depairing influences the proximity-induced pairing and results in a
quadratic decrease of the induced superconducting gap:
$\Delta(B)=\Delta\,(1-B^2/B_*^2)$, where $B_*\sim \Phi_0/A$ and $A$ is the
cross-section of the nanowire.
A simple model which includes both orbital and Zeeman effect, but no spin-orbit
coupling, yields $B_* \approx 400\,$mT when fitted to the experimental data (see
Supplementary Material for details).
In this case, the fit is insensitive to the value of the $g$-factor.
However, the model also predicts the occurrence, at $\varphi=\pi$, of a
fermion-parity switch at a field $B_\textrm{sw}<B_*$ whose value depends on the
$g$-factor.
Because agreement with the experimental data imposes the condition that
$B_\textrm{sw}>300\,$mT, in the Supplementary Material we show that this
scenario requires $g\lesssim 5$, which is lower than $g$-factor values measured earlier
in InAs nanowire channels \citeM{das_2012,
higginbotham_natphys_2015,albrecht_nature_2016}.

Furthermore, we can consider the qualitative effect of the inclusion of a weak
spin-orbit coupling ($E_\textrm{SO}\ll \Delta$) in this model containing only
the orbital and Zeeman effects.
We note that, without spin-orbit coupling, the upper Andreev level
$E^+_\textrm{ABS}(B)$ crosses a continuum of states $\Delta(B)$ with opposite
spin upon increasing the magnetic field (see Fig.~S11c in the Supplementary
Material). The crossing happens at a field of $B_\textrm{cross}$ whose value
depends on the $g$-factor: using the upper bound for $g$ derived in the last
paragraph, $g\approx 5$, we can estimate $B_\textrm{cross}\approx150\,$mT.
At this magnetic field, a weak spin-orbit coupling results in an avoided
crossing between the Andreev level $E^+_\textrm{ABS}(B)$ and the continuum.
As a consequence, when $B>B_\textrm{cross}$, the energy $E^+_\textrm{ABS}(B)$ is
bounded by the edge of the continuum and it is markedly lower than its value in
the absence of spin-orbit coupling.
In turn, this results in a decrease of the transition energy $E_\textrm{tot}(B)$
at $B>B_\textrm{cross}$, to the extent that such a model containing the joint
effect of orbital depairing and weak spin-orbit coupling would depart from the
experimental data in the range $150\,$mT $< B < 300\,$mT (see dotted line in
Fig.~4b).
Thus, although based on the geometry of the experiment we cannot rule out the
presence of an orbital effect of the magnetic field, these considerations imply
that it does not play a dominant role in the quadratic suppression of the
transition energy in the present measurements.

We finally note that in all cases we neglect the effect of $B$ on the Al thin
film, justified by its in-plane critical magnetic field exceeding $2\,$T
\citeM{meservey1971}.

\begin{figure}
\centering
\includegraphics[width=0.5\textwidth]{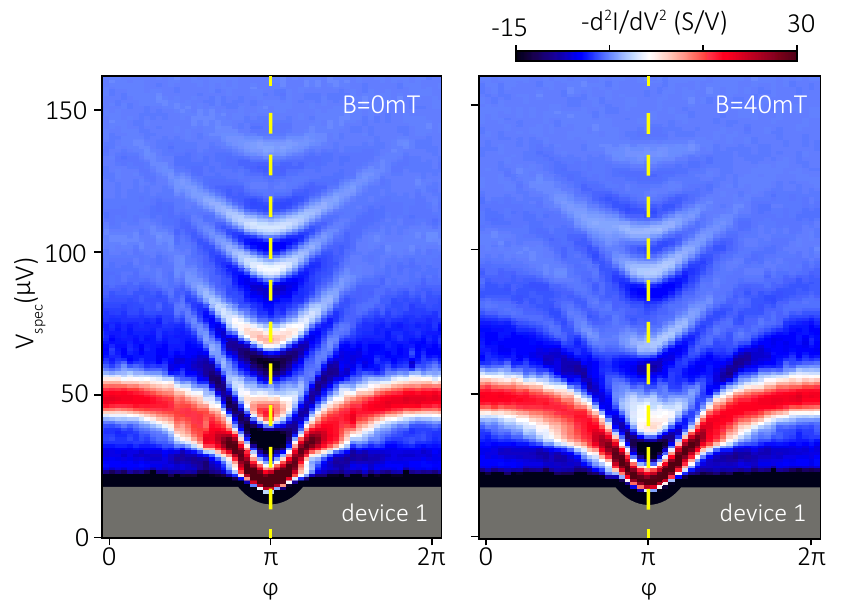}
\caption{\textbf{Time-reversal symmetry-broken ABS in magnetic field.} The symmetry axis at
$\varphi=\pi$ at zero magnetic field is denoted by yellow dashed line. Note that
at $B=40\,$mT the observed spectrum does not obey the mirror symmetry with
respect to the same line. The data was taken on device 1 at
$V_g=-20\,$mV. Grey regions denote lack of data due to bias instability of the
circuit.}
\end{figure}

We present the ABS spectrum in the presence of several transport channels in
Fig.~5.
While at zero magnetic field (left panel) the data is symmetric around
$\varphi=\pi$, in a finite magnetic field (right panel) the data exhibits an
asymmetric flux dependence (see the yellow dashed line as a guide to the eye).
This should be contrasted with Fig.~4a where the data for a single-channel wire
are presented at different values of the magnetic field: each of the traces is
symmetric around $\varphi=\pi$.
This behavior agrees with theoretical calculations in the short-junction limit,
which show that this asymmetry can arise in a Josephson junction with broken
time-reversal and spin-rotation symmetries as well as more than one transport
channel \citeM{doi:10.7566/JPSJ.82.054703}.
While the data is asymmetric with respect to $\varphi=\pi$, there is no visible
shift of the local energy minima away from this point.
This observation is consistent with the absence of an anomalous Josephson
current \citeM{Krive2004,PhysRevLett.101.107005,2015arXiv151201234S} for our
specific field configuration (magnetic field parallel to the wire), in agreement
with theoretical expectations
\citeM{PhysRevB.82.125305,PhysRevB.93.155406,PhysRevB.92.125443}.

In conclusion, we have presented microwave spectroscopy of Andreev bound states
in semiconductor channels where the conductive modes are tuned by electrostatic
gates and we have demonstrated the effect of Zeeman splitting and spin-orbit
coupling. The microwave spectroscopy measurements shown here could provide a new
tool for quantitative studies of Majorana bound states, complementing
quasiparticle tunneling experiments \citeM{Mourik1003, das_2012}. Furthermore, we
have provided direct evidence for the time-reversal symmetry breaking of the
Andreev bound state spectrum in a multichannel ballistic system. This result
paves the way to novel Josephson circuits, where the critical current depends on
the current direction, leading to supercurrent rectification effects
\citeM{Villegas1188, PhysRevLett.101.107001} tuned by electrostatic gates.

\section{Corresponding author}

Correspondence and request of materials should be sent to Attila Geresdi.

\section{Data availability}

The datasets generated and analysed during this study are available at the
4TU.ResearchData repository, DOI:
10.4121/uuid:8c4a0604-ac00-4164-a37a-dad8b9d2f580 (Ref.
\citeM{rawdata}).

\section{Acknowledgements}

The authors thank L.~Bretheau, \c{C}.~\"O.~Girit, L.~DiCarlo, M.~P.~Nowak and
A.~R.~Akhmerov for fruitful discussions, and R.~van Gulik, T.~Kriv\'achy,
A.~Bruno, N.~de Jong, J.~D.~Watson, M.~C.~Cassidy, R.~N.~Schouten and
T.~S.~Jespersen for assistance with fabrication and experiments. This work has
been supported by the Danish National Research Foundation, the Villum
Foundation, the Dutch Organization for Fundamental Research on Matter (FOM), the
Netherlands Organization for Scientific Research (NWO) by a Veni grant,
Microsoft Corporation Station Q and a Synergy Grant of the European Research
Council. B.~v.~H.~was supported by ONR Grant Q00704. L.~I.~G.~and
J.~I.~V.~acknowledge the support by NSF Grant DMR-1603243.

\section{Author contributions} 

D.~J.~v.~W., A.~P.~and D.~B.~performed the
experiments. B.~v.~H., J.~I.~V.~and L.~I.~G.~developed the theory to analyze the
data. P.~K.~and J.~N.~contributed to the nanowire growth.
D.~J.~v.~W., A.~P.~and D.~B.~fabricated the samples.
L.~P.~K.~and A.~G.~designed and supervised the experiments. D.~J.~v.~W.,
B.~v.~H., L.~P.~K.~and A.~G.~analyzed the data. The manuscript has been prepared
with contributions from all the authors.

\bibliographystyleM{apsrev4-1}
\bibliographyM{references_ABS_nourl}

\onecolumngrid
\pagebreak

\begin{center}
\textbf{\large Supplementary online material} 

\textbf{\large Microwave spectroscopy of spinful Andreev bound states in ballistic
semiconductor Josephson junctions}
\end{center}

\setcounter{figure}{0}
\makeatletter 
\renewcommand{\thefigure}{S\@arabic\c@figure}
\makeatother

\setcounter{equation}{0}
\makeatletter 
\renewcommand{\theequation}{S\@arabic\c@equation}
\makeatother

\setcounter{table}{0}
\makeatletter 
\renewcommand{\thetable}{S\@arabic\c@table}
\makeatother

\setcounter{page}{1}
 
\section{Device fabrication}

The devices are fabricated on commercially available undoped Si wafers with a
$285\,$nm thick thermally grown SiO$_x$ layer using positive tone electron beam
lithography. First, the electrostatic gates and the lower plane of the coupling
capacitors are defined and Ti/Au ($5\,$nm/$15\,$nm) is deposited in a
high-vacuum electron-beam evaporation chamber. Next, the decoupling resistors
are created using Cr/Pt ($5\,$nm/$25\,$nm) with a track width of $100\,$nm,
resulting in a characteristic resistance of $100\,\Omega/\mu$m. Then, a $30\,$nm
thick SiN$_x$ layer is sputtered and patterned to form the insulation for the
coupling capacitors and the gates. We infer $C_c=400\,$fF based on the surface
area of $6.5\times 30\,\mu$m$^2$ and a typical dielectric constant
$\varepsilon_r=7$.

In the following step, the tunnel junctions
are created using the Dolan bridge technique by depositing $9$ and $11\,$nm
thick layers of Al with an intermediate oxidization step \emph{in-situ} at
$1.4\,$mbar for 8 minutes.
Then, the top plane of the coupling capacitors is defined and evaporated (Ti/Au,
$20\,$nm/$100\,$nm) after an \emph{in-situ} Ar milling step to enable metallic
contact to the Al layers. Next, the InAs nanowire is deterministically deposited
with a micro-manipulator on the gate pattern \citeS{suppFlohr_2011}.

The channel of device 1 is defined by wet chemical etch of the aluminium shell
using Transene D at $54\,^\circ$C for $12\,$seconds. The channel
of device 2 is determined by \emph{in-situ} patterning, where an adjacent
nanowire casted a shadow during the epitaxial deposition of aluminium
\citeS{suppKrogstrup_shadow}. The superconducting layer thickness was approximately
$10\,$nm for both devices deposited on two facets.

Finally, the nanowire is contacted to the rest of the circuit by performing Ar
plasma milling and subsequent NbTiN sputter deposition to form the loop of the
hybrid SQUID. We show the design parameters of the devices in Table S1.

\begin{table}[!ht]
\begin{tabular}{|c||c|c|}
\hline
 & Device 1 & Device 2 \\ \hline \hline
 Channel length (nm) & 100 & 40 \\ \hline
 Tunnel junction area (nm$^2$) & $400\times120$ & $200\times120$ \\ \hline
 Flux periodicity ($\mu$T) & 38 & 120 \\ \hline
 Spectrometer junction area (nm$^2$) & $120\times120$ & $120\times120$ \\ \hline
\end{tabular}
\caption{\textbf{Geometry of the devices featured in the current study.}}
\end{table}

\newpage

\section{Measurement setup}

\begin{figure}[H]
\centering
\includegraphics[width=0.75\textwidth]{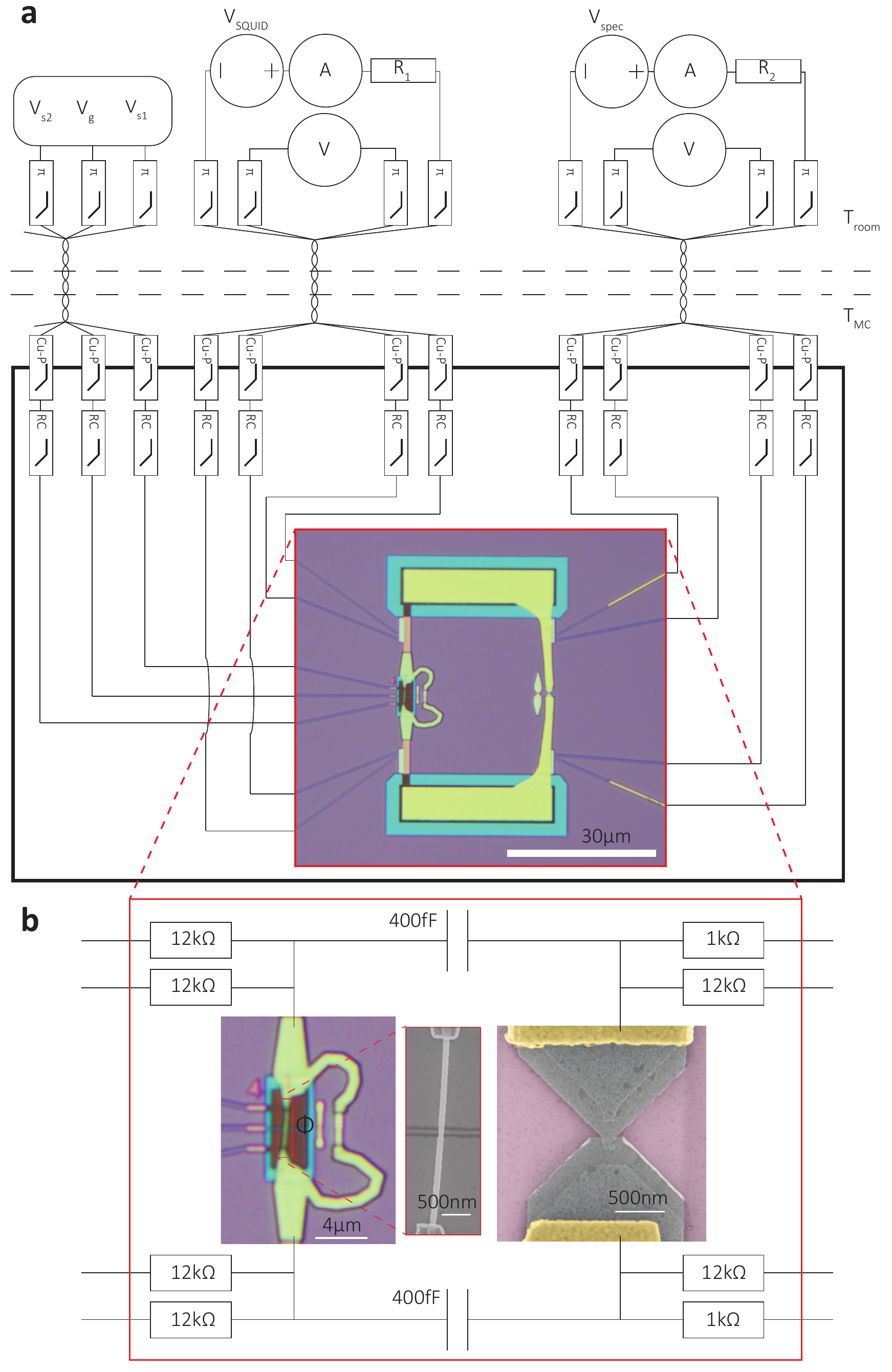}
\caption{\textbf{Detailed schematics of the measurement setup.} The inset of
panel (a) shows a bright field optical image of device 1. The solid black box
denotes the radiation shielded environment thermally anchored to $12\,$mK.
(b) On-chip lumped circuit elements attached to the hybrid SQUID (on the left)
and the spectrometer Josephson junction (on the right).}
\end{figure}

The measurements were performed in a Leiden Cryogenics CF-1200 dry dilution
refrigerator with a base temperature of $12\,$mK equipped with Cu/Ni shielded
twisted pair cables thermally anchored at all stages of the refrigerator to
facilitate thermalization. Noise filtering is performed by a set of $\pi$-LC
filters ($\sim 100\,$MHz) at room temperature and copper-powder filters ($\sim
1\,$GHz) in combination with two-pole RC filters ($\sim 100\,$kHz) at base
temperature for each measurement line. The schematics of the setup
is shown in Fig.~S1.

\section{Device circuit parameters}

We characterise the circuit based on the plasma resonance observed with the
semiconductor nanowire gated to zero conductance, i.e.~full depletion. In this
regime, we infer the environmental impedance $\Re[Z(\omega)]$ based on Eq.~(2)
in the main text and assume the following form, which is valid for a parallel
LCR circuit:
\begin{equation}
\Re[Z(x)]=\frac{Z_0 Q}{1+\frac{Q^2}{x^2}(1-x^2)^2},
\end{equation}
with $x=\omega/\omega_0$ the dimensionless frequency. The resonance of the
circuit is centered at $\omega_0=(LC)^{-1/2}$ with a quality factor of
$Q=R\sqrt{\frac{C}{L}}$ and a characteristic impedance of $Z_0=\sqrt{L/C}$.
Consistently with this single mode circuit, we find one peak in the $I(V)$ trace
of the spectrometer that we fit to Eq.~(S1) (Fig.~S2). We find a good
quantitative agreement near the resonance peak, however the theoretical curve consistently
deviates at higher voltages, i.e.~higher frequencies. We attribute this
discrepancy to additional losses or other resonant modes of the circuit not
accounted for by Eq.~(S1).

In addition, we use the superconducting gap and the linear resistance of the
junctions to determine the Josephson energy $E_J$ and the Josephson
inductance $L_J$. With these, we infer the circuit parameters listed in Table
S2.

\begin{figure}[H]
\centering
\includegraphics[width=\textwidth]{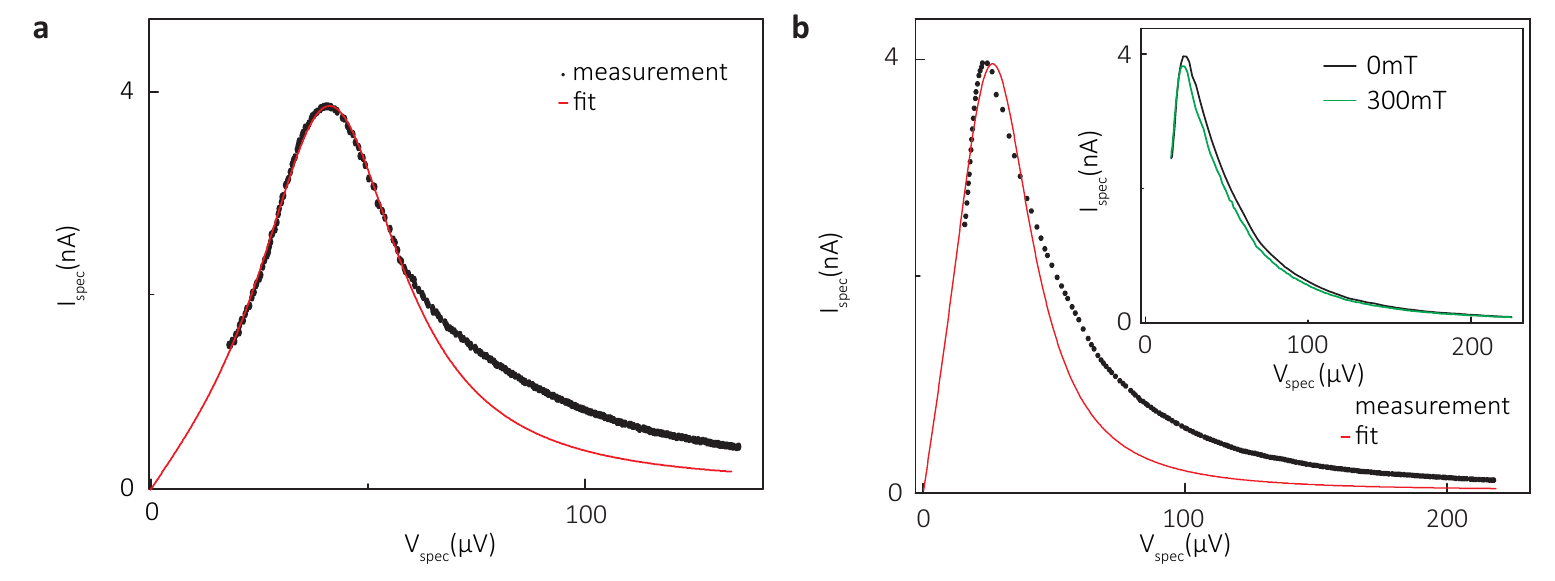}
\caption{\textbf{Plasma resonance of the circuit.} The measured (black dots) and
fitted (solid red line) $I(V)$ trace of the spectrometer junction for device 1
(a) and for device 2 (b) respectively, with the nanowire in full depletion.
The fits are based on Eq.~(S1), see text. Note that we omitted the supercurrent
branch for clarity. In panel (b), the inset shows the spectrometer response to
an in-plane magnetic field of $300\,$mT.}
\end{figure}

\begin{figure}[H]
\centering
\includegraphics[width=0.85\textwidth]{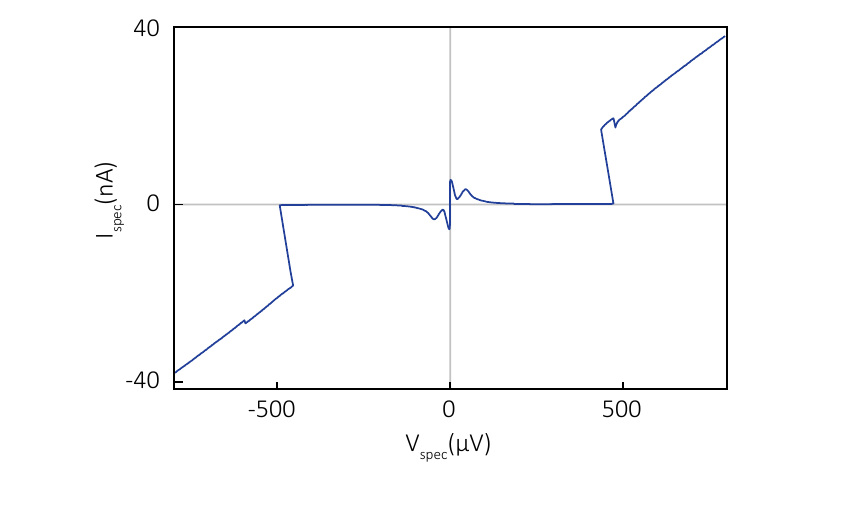}
\caption{\textbf{Large scale I(V) trace of the spectrometer junction.} The
subgap features are shown in detail in Fig.~S2a. The back-bending near
$eV_\textrm{spec}\approx 2\Delta_\textrm{spec} = 482\,\mu$eV is attributed to a
local overheating of the junction due to a large quasiparticle current density
above the gap edge. The data was taken on device 1.}
\end{figure}

\begin{table}[!ht]
\onecolumngrid
\begin{tabular}{|c||c|c|c|c|}
 \hline
 & Device 1 & Device 2 \\ \hline \hline
 Tunnel junction resistance $R_J$ (k$\Omega$) & 4.80  & 10.7 \\ \hline
 Tunnel junction gap $\Delta_J$ ($\mu$eV) & 245 & 250 \\ \hline
 Tunnel junction critical current $I_\textrm{c,J}=\frac{\pi\Delta_J}{2eR_J}$
 (nA) & 80.2 & 36.7\\
 \hline $E_J=\frac{\hbar I_\textrm{c,J}}{2e}$ ($\mu$eV) & 165 & 75.5 \\
 \hline
 Tunnel junction inductance $L_J=\frac{\Phi_0}{2\pi I_\textrm{c,JJ}}$ (nH) &
 4.10 & 8.94 
 \\
 \hline Spectrometer resistance $R_\textrm{spec}$ (k$\Omega$) & 17.1 & 18.4 \\
 \hline Spectrometer gap $\Delta_\textrm{spec}$ ($\mu$eV) & 241 & 249 \\ \hline
 Spectrometer critical current $I_\textrm{c,spec}=\frac{\pi\Delta_\textrm{spec}}{2eR_\textrm{spec}}$
 (nA) & 22.2 & 21.3 \\ \hline
 Shunt resistance $R$ ($\Omega$) & 634 & 743 \\ \hline
 Shunt capacitance $C$ (fF) & 12.6 & 11.1 \\ \hline
 Charging energy $E_c=\frac{2e^2}{C}$ ($\mu$eV) & 25.44 & 29.1 \\ \hline
 Plasma frequency $f_p=\frac{1}{2\pi\sqrt{L_JC}}$ (GHz) &
 22.9 & 16.0 \\ \hline
 Characteristic impedance $Z_0=\sqrt{\frac{L_J}{C}}$ ($\Omega$) & 551 & 897
 \\
 \hline
 Quality factor $Q=R\sqrt{\frac{C}{L_J}}$ & 1.15 & 0.83
 \\
  \hline
\end{tabular}
\caption{\textbf{Circuit parameters of the devices featured in the current
study.}}
\end{table}

\clearpage

\newpage
\section{Additional datasets}

\subsection{Spectrum analysis}

Peaks in the $I(V)$ trace of the spectrometer correspond to peaks in
$\Re[Z(\omega)]$, i.e.~allowed transitions of the environment coupled to the
spectrometer. In order to remove the smooth background of the plasma mode (see
Fig.~S2), we evaluate $-d^2 I/dV^2(V)$, the second derivative of the $I(V)$ to
find peaks in $\Re[Z(\omega)]$ after applying a Gaussian low pass filter with
standard deviation of $1.5\,\mu$V. We benchmark this method in Fig.~S4, and find
that the peaks where $-d^2 I/dV^2(V) > 0$ correspond to the peaks in $I(V)$ and
hence $-d^2 I/dV^2(V)$ is a good measure of the transitions detected by the
spectrometer junction. 

\begin{figure}[H]
\centering
\includegraphics[width=\textwidth]{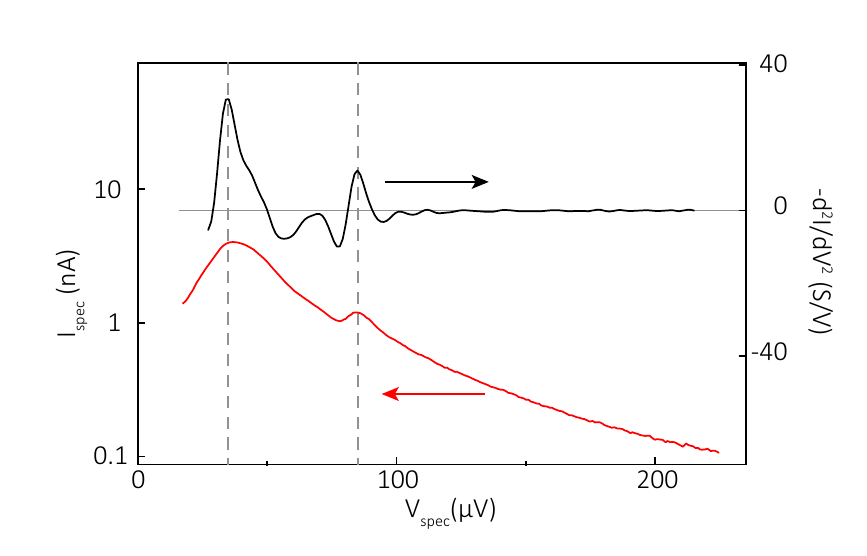}
\caption{\textbf{Spectrum analysis by second derivative.} The $I(V)$ (red line,
left axis) and the corresponding $-d^2I/dV^2(V)$ trace (black line, right axis)
of the spectrometer showing the same peaks denoted by dashed lines. Note that only
peaks above $-d^2I/dV^2(V)=0$ (grey horizontal line) correspond to actual
transitions. This dataset was taken on device 1, at $V_g=-1410\,$mV, phase
biased to $\varphi=\pi$.}
\end{figure}

Alternatively, the background can be removed by linewise subtracting the
detector response at $\varphi=0$ \citeS{suppbretheau_nature}, where the ABS does not
contribute to the spectrometer response \citeS{suppPhysRevB.87.174521}. We show the
result of this analysis in Fig.~S5. Notably, the phase dependence of the plasma
mode gives rise to additional features near $\varphi=\pi$. Furthermore, datasets
exhibiting hybridization between the ABS and plasma mode cannot be evaluated
by this method. However, the line subtraction and the second derivative are in
agreement if there is sufficient spacing between the plasma mode and the
ABS line (see Fig.~2b and Fig.~S5 for comparison).

\begin{figure}[H]
\centering
\includegraphics[width=\textwidth]{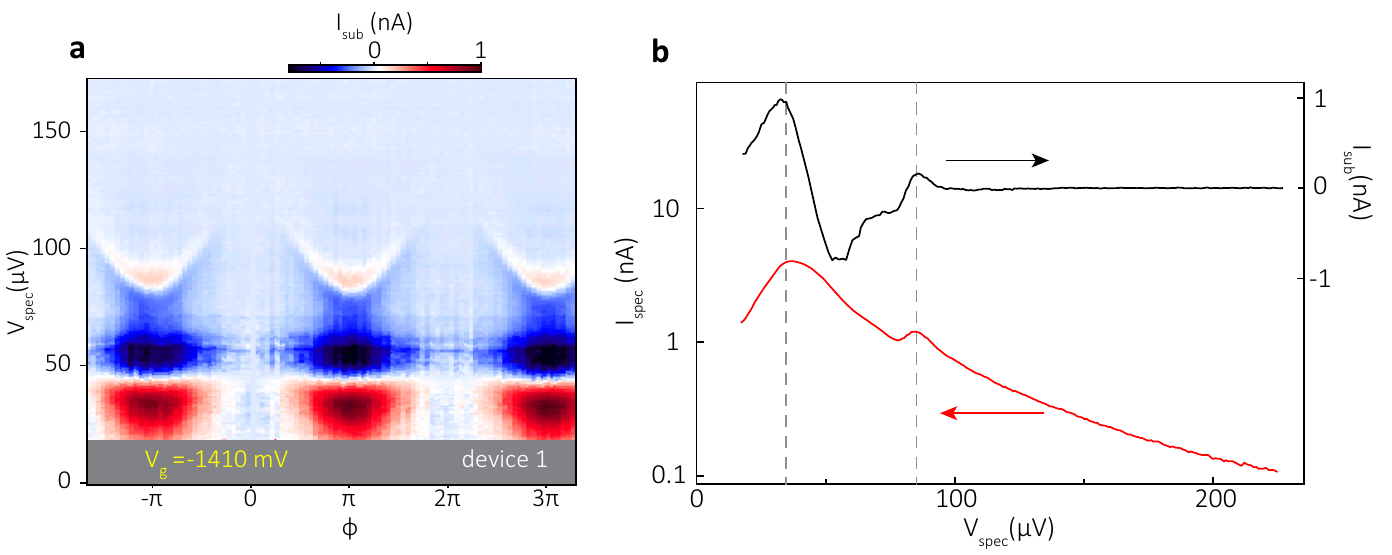}
\caption{\textbf{Spectrum analysis by background subtraction.} (a) 
$I_\textrm{sub}(\varphi)=I_\textrm{spec}(\varphi)-I_\textrm{spec}(\varphi=0)$
spectrometer current after subtracting the line trace at $\phi=0$. (b) Single
linetrace of the raw data $I_\textrm{spec}(\varphi=\pi)$ (red line, left axis)
and $I_\textrm{sub}(\varphi=\pi)$ (black line, right axis). This dataset was taken
on device 1, at $V_g=-1410\,$mV.}
\end{figure}

\newpage
\subsection{I(V) trace of the hybrid SQUID}

We measure the $I(V)$ trace of the hybrid SQUID as a function of the gate
voltage $V_g$ at $V_\textrm{spec}=0$ (Fig.~S6) and find that the subgap
conductance increases with increasing gate voltage, in qualitative agreement
with the contribution of multiple Andreev reflection (MAR). The zero voltage
data corresponds to the supercurrent branch and the dashed lines denote the
bias range where there is no data due to the bias instability of the driving
circuit. In addition, we find a back-bending at the gap edge
$eV_\textrm{SQUID}=2\Delta_J$, attributed to self-heating effects in the tunnel
junction.

We evaluate $G$ in Fig.~2e of the main text in the bias voltage range
$-V_\textrm{SQUID}=350\ldots 430\,\mu\textrm{V}>2\Delta$. We note that due to
the soft superconducting gap in the nanowire junction, we did not identify MAR
features after subtracting the current background of the tunnel junction. 

\begin{figure}[H]
\centering
\includegraphics[width=0.85\textwidth]{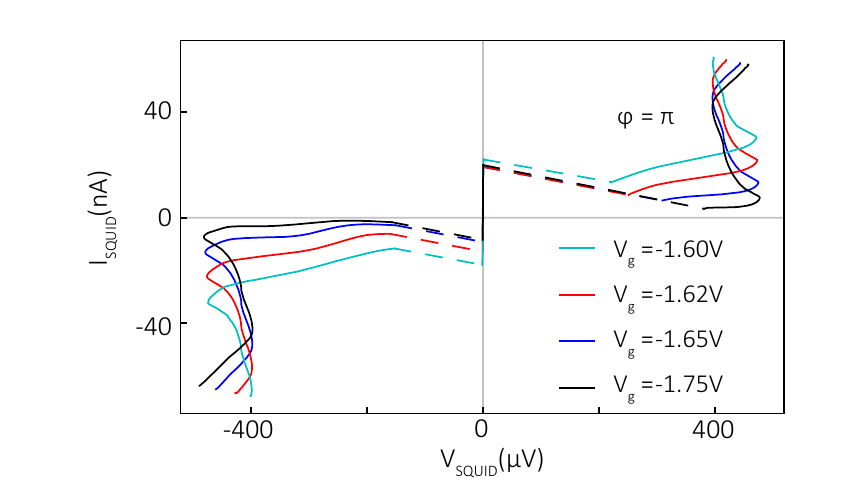}
\caption{\textbf{The $I(V)$ trace of the hybrid SQUID.} At $V_g=-1.75\,$V, the
nanowire is in full depletion, thus the corresponding $I(V)$ trace represents
the Al/AlO$_x$/Al tunnel junction in the hybrid SQUID. The bias
voltage $V_\textrm{SQUID}$ was swept from the left to the right. The data was
taken on device 1.}
\end{figure}

\newpage
\subsection{Fit of ABS with high transmission}

\begin{figure}[H]
\centering
\includegraphics[width=\textwidth]{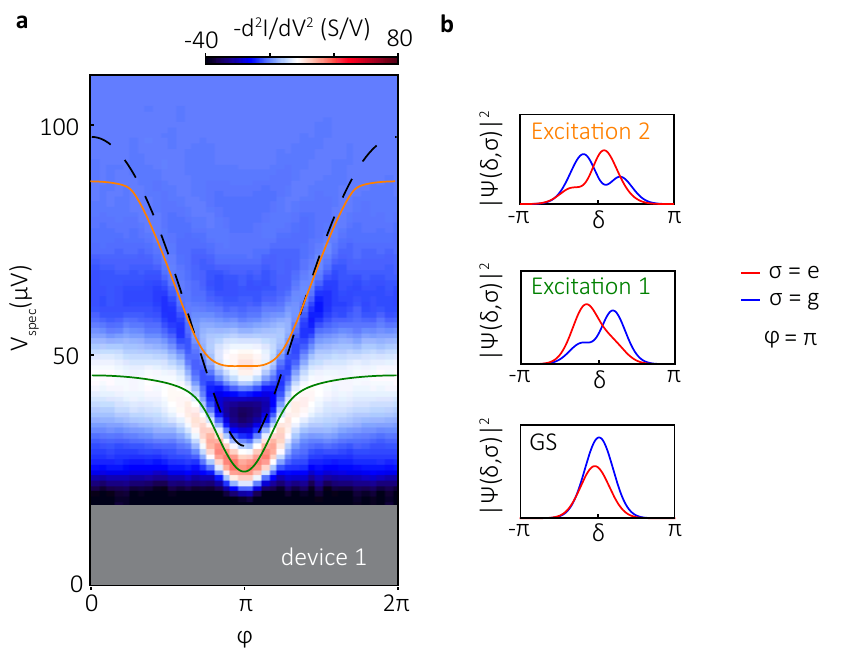}
\caption{\textbf{Experimental data and fit to the theory for ABS with high
transmission.} In this figure we show the numerical fit to the model of Eq.~(3),
similarly to Fig.~3a, but for a different dataset taken at $V_g=-1.525\,$V on
device 1. The figure shows that the model of Eq.~(3) can accurately predict the
avoided crossing originating in the coupling between the ABS and the plasma mode.
Best-fit parameters are $\Delta=97.5\pm1.7\,\mu$eV and $T=0.90\pm0.01$. Dashed
line denotes the undressed Andreev level defined by Eq.~(1) in the main text. We
note that the extracted value for $\Delta$ is lower than in Fig.~3a.
This may stem from the fit underestimating the gap, since most of the datapoints
are around $\varphi= \pi$, or due to a genuine dependence of $\Delta$ on $V_g$
because of the change in the wavefunction overlap as a result of the
electrostatic gating \protect\citeS{supp1367-2630-18-3-033013}.
In panel (b), we show the probability density for the ground state (GS) and the two
observed excited states denoted by the green and orange lines,
respectively in panel (a) at $\varphi=\pi$.}
\end{figure}

\newpage
\subsection{Time-reversal symmetry-broken ABS in bipolar magnetic field}
\begin{figure}[H]
\centering
\includegraphics[width=\textwidth]{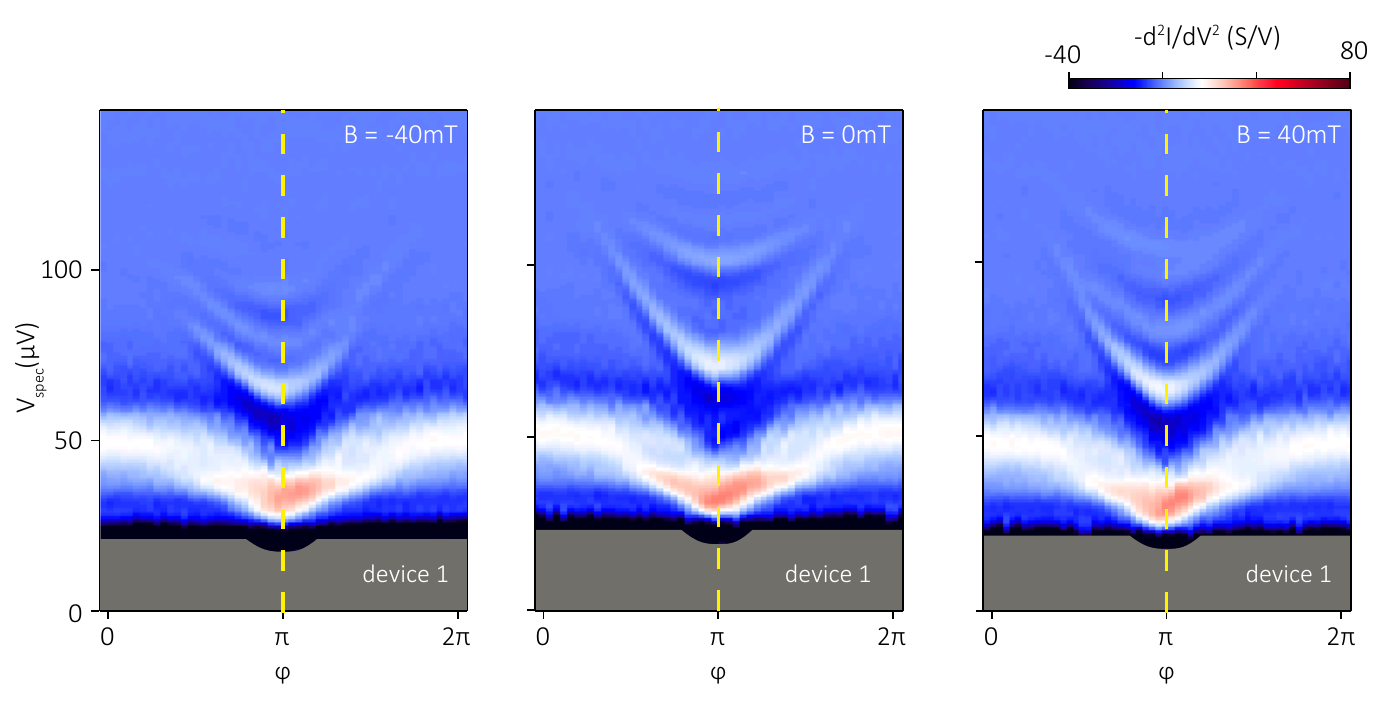}
\caption{\textbf{Symmetry-broken ABS in a bipolar magnetic field.} The full
spectrum is symmetric around $\varphi=\pi$ at zero magnetic field (center
panel) with the mirror axis denoted by the yellow dashed line. Note the
asymmetry of the two lowermost ABS transitions at $B=\pm40\,$mT. The antisymmetric contribution is
most visible at $V_\textrm{spec}\approx100\,\mu$V, which develops an opposite
shift for positive and negative magnetic fields, respectively. The data
was taken on device 1 at $V_g=-770\,$mV. Grey regions denote lack of data due to
bias instability of the circuit.}
\end{figure}

\newpage
\section{Theory}

\subsection{Estimate of the ABS-plasma resonance avoided crossing}

Before describing the quantum model of the circuit in detail, we discuss the
estimate for the energy splitting at the avoided crossing between the ABS
transition and the plasma frequency shown in Fig.~2d.

For simplicity, we model the plasma oscillations as a bosonic mode with a
flux-independent frequency given by $\hbar \omega_p = \sqrt{2 E_J E_C}$, and the
weak link as a two-level system, with energies $\pm E_\textrm{ABS}(\varphi)$
defined by Eq.~(1) in the main text.
This system with the two independent degrees of freedom is described by the
Hamiltonian $\hat{H}_0 = \hbar\omega_p(\hat{a}^\dagger \hat{a}
+\tfrac{1}{2})+E_\textrm{ABS}\,\hat{\sigma}_3$. Next, we add the coupling term
corresponding to the excitation of the weak link due to the voltage oscillations
induced by the junction in the form
\begin{equation}
H_g(\varphi) = g(\varphi)\,\sqrt{z}\,(\hat{a}^\dagger+\hat{a})\,\hat{\sigma}_1.
\end{equation}
where $z=\sqrt{E_C/2E_J}$. This term describes a linear coupling between the
two-level system and the phase difference across the junction. $g(\varphi)$ is
then given by the current matrix element between the ground and excited states
of the weak link, which was derived in Ref.~\citeS{suppPhysRevB.87.174521}:
\begin{equation}\label{eq:matrixelement}
g(\varphi) = \Delta\,T
\sqrt{1-T}\,\sin^2(\varphi/2)\frac{\Delta}{E_\textrm{ABS}(\varphi)}\,.
\end{equation}
The square of this current matrix element gives the microwave absorption rate of
the weak link,  plotted in Fig.~3c (black dashed line) of the main text.
From the coupling Hamiltonian, we immediately obtain that at $\varphi=\pi$, the
splitting is
\begin{equation}
\varepsilon = \Delta\,T\,\sqrt{z}
\end{equation}
which is the expression used for the estimate in the main text. We note that
Eq.~(S4) is the lowest-order estimate of the avoided crossing in the
small parameter $\sqrt{z}$. The relatively high value
$\sqrt{z}\approx 0.52$ of device 1 may explain the discrepancy between this
simple estimate and the observed value, which is captured by the
full model, see Fig. S7. 
Finally, we note that the expression~(\ref{eq:matrixelement}) was also derived
in Ref.~\citeS{suppPhysRevB.90.134506} starting from the full model (see next section).
In particular, the quantity $\Omega_x(\varphi)$ in
Ref.~\citeS{suppPhysRevB.90.134506} is equal to $\sqrt{z}\,g(\varphi)$.

\subsection{Hamiltonian description of the hybrid SQUID}

We now describe the theoretical model of the hybrid SQUID that was used to fit
the experimental data. Our model is based on
Refs.~\citeS{suppPhysRevLett.90.087003} and~\citeS{suppPhysRevB.71.214505}. The
Hamiltonian of the model is Eq.~(3) of the main text, repeated here for
convenience:
\begin{equation}\label{eq:H_squid}
\hat{H} = E_C \hat{N}^2 + E_J(1-\cos\hat{\delta}) +
\hat{H}_\textrm{ABS}(\varphi-\hat{\delta})\,,
\end{equation}
with $[\hat{\delta}, \hat{N}]=i$. The Hamiltonian of the weak link is~\citeS{suppPhysRevLett.90.087003}
\begin{equation}
\hat{H}_\textrm{ABS}(\phi)=\Delta\,\hat{U}(\phi)\,\left[\,\cos(\phi/2)\,\hat{\sigma}_3+\sqrt{1-T}\,\sin(\phi/2)\,\hat\sigma_2\right]\,\hat{U}^\dagger(\phi)\,,\\
\end{equation}
with $\hat{U}(\phi) = \exp\,(-i\,\sqrt{1-T}\,\hat\sigma_1\,\phi/4)$. Here
$\hat{\sigma}_2$ and $\hat{\sigma}_3$ are two Pauli matrices which act on a
space formed by the ground state of the weak link and an excited state with a
pair of quasiparticles in the weak link. By expanding the product above, the
Hamiltonian can be put in the form 
$\hat{H}_\textrm{ABS}(\phi) = V_2(\phi)\,\hat\sigma_2+V_3(\phi)\,\hat\sigma_3$.
The two functions $V_2$ and $V_3$ are:
\begin{align}
V_2(\phi)&=\Delta\,\sqrt{1-T} \sin \left(\phi/2\right) \cos \left(\sqrt{1-T}\phi/2\right)- \Delta\,\cos\left(\phi/2\right) \sin \left(\sqrt{1-T}\phi/2\right)\,,\\
V_3(\phi)&=\Delta\,\sqrt{1-T} \sin \left(\phi/2\right) \sin \left(\sqrt{1-T}\phi/2 \right)+\Delta\,\cos \left(\phi/2\right) \cos \left(\sqrt{1-T} \phi/2 \right)\,,
\end{align}
We introduce the ground ($\ket{g}$) and excited states ($\ket{e}$) of the weak
link in the presence of an equilibrium phase difference, 
\begin{subequations}
\begin{align}
\hat{H}_\textrm{ABS}(\phi)\,\ket{g} &= - E_\textrm{ABS}(\phi) \ket{g}\,,\\
\hat{H}_\textrm{ABS}(\phi)\,\ket{e} &= + E_\textrm{ABS}(\phi) \ket{e}\,,
\end{align}
\end{subequations}
where $E_\textrm{ABS}(\phi)$ is given in Eq.~(1) of the main text. In the basis
$\ket{\pm}$ of eigenstates of $\hat{\sigma}_3$,
$\hat\sigma_3\,\ket{\pm}=\pm\,\ket{\pm}$, they are given by
\begin{subequations}\label{eq:g,e}
\begin{align}
\ket{g} &= c_{g+}(\phi)\,\ket{+} + c_{g-}(\phi)\,\ket{-}\,,\\
\ket{e} &= c_{e+}(\phi)\,\ket{+} + c_{e-}(\phi)\,\ket{-}\,,
\end{align}
\end{subequations}
with the coefficients 
\begin{subequations}
\begin{align}
c_{g+}(\phi) &= i\,\frac{E_A(\phi)-V_3(\phi)}{\sqrt{2E_A(\phi)[E_A(\phi)-V_3(\phi)]}}\,,\;\;&c_{g-}(\phi) &= \frac{V_2(\phi)}{\sqrt{2E_A(\phi)[E_A(\phi)-V_3(\phi)]}}\,,\\
c_{e+}(\phi) &= -i\,\frac{E_A(\phi)+V_3(\phi)}{\sqrt{2E_A(\phi)[E_A(\phi)-V_3(\phi)]}}\,,\;\;&c_{e-}(\phi) &= \frac{V_2(\phi)}{\sqrt{2E_A(\phi)[E_A(\phi)+V_3(\phi)]}}\,.
\end{align}
\end{subequations}
The coefficients are normalized: 
\begin{equation}
|c_{g+}(\phi)|^2+|c_{g-}(\phi)|^2=|c_{e+}(\phi)|^2+|c_{e-}(\phi)|^2=1\,.
\end{equation}

To find the resonant frequencies of the hybrid SQUID, we solve the eigenvalue
problem ${\hat{H}\ket{\Psi} = E\ket{\Psi}}$ numerically. We adopt the basis
$\ket{\delta, \pm}\equiv\ket{\delta}\otimes\ket{\pm}$ for the joint eigenstates
of the $\hat\delta$ and $\hat\sigma_3$ operators:
$\hat\delta\,\hat\sigma_3\,\ket{\delta,\pm} =
(\hat\delta\,\ket{\delta})\otimes(\hat\sigma_3\,\ket{\pm}) =
\pm\,\delta\,\ket{\delta,\pm}$. For the numerical solution, we use a truncated Hilbert
space where the phase interval $[-\pi,\pi)$ is restricted to $M$ discrete
points, with lattice spacing $2\pi/M$. A complete basis of the truncated Hilbert
space is given by the $2M$ vectors $\ket{\delta_k}\otimes\ket{\pm}$ with
$\delta_k = 2\pi k /M$ ($k= 0, \pm 1, \pm 2, \dots, \pm (M-1)/2$), and
$\ket{\pm}$ the eigenvector of $\hat{\sigma}_3$. The Hamiltonian is thus
represented as a $2M\times 2M$ matrix in this basis and diagonalized
numerically. We choose the parameter $M$ large enough to guarantee convergence
of the eigenvalues.

Once the spectrum is known, we use the transition frequencies from the ground
state, $\omega_n = E_n - E_\textrm{GS}$, to do a least-square fit to the
experimental data.
The details of the numerical procedure are listed in the Jupyter
notebooks available at \citeS{supprawdata}.

Once an eigenstate $\ket{\Psi}$ is determined numerically, we 
represent its two-component wavefunction in the basis of the weak link
eigenstates $\{\ket{g}, \ket{e}\}$ from Eq.~\eqref{eq:g,e}, evaluated at
$\phi=\varphi$:
\begin{equation}
\ket{\Psi} = \sum_{\delta}\,\sum_{\sigma = g,e}\, \Psi(\delta, \sigma) \ket{\delta, \sigma}\;,\;\;\Psi(\delta, \sigma) = \braket{\delta, \sigma}{\Psi}\,,
\end{equation}
where 
\begin{equation}
\ket{\delta,\sigma} = \,\ket{\delta}\,\otimes\,(c_{\sigma+}(\varphi)\,\ket{+} + c_{\sigma-}(\varphi)\,\ket{-})\,.
\end{equation}

The probability densities $|\Psi(\delta,\sigma)|^2$ plotted in Fig.~3b and
Fig.~S7b allow us to evaluate at a glance whether the eigenstate $\ket{\Psi}$
has a large overlap with the excited state $\sigma=\ket{e}$ of the (decoupled)
weak link.

Finally, in Fig.~3c we show the numerical prediction for
the visibility of the ABS transition as a function of the phase bias, $\phi$.
The visibility is determined by the absolute square of current operator matrix element
$\bra{GS}\,\hat{J}(\varphi)\,\ket{\Psi}$ between the ground state $\ket{GS}$ and
the excited state $\ket{\Psi}$ of $\hat{H}$ corresponding to the ABS transition.
The current operator is~\citeS{suppPhysRevB.71.214505}
\begin{equation}
\hat{J}(\varphi) = E_J\,\sin(\hat\delta) + \frac{\pd H_\textrm{ABS}(\varphi-\hat\delta)}{\pd\,\hat\delta}\,.
\end{equation}

\subsection{Equilibrium phase drop}

In the main text, we have often assumed that
the equilibrium phase drop across the weak link, $\phi$, is close to the total
applied phase, $\phi \approx \varphi$. Here, we verify this
assumption by calculating the equilibrium phase drop of the hybrid
SQUID model we presented in the previous section. 

\begin{figure}
\centering
\includegraphics[width=0.8\textwidth]{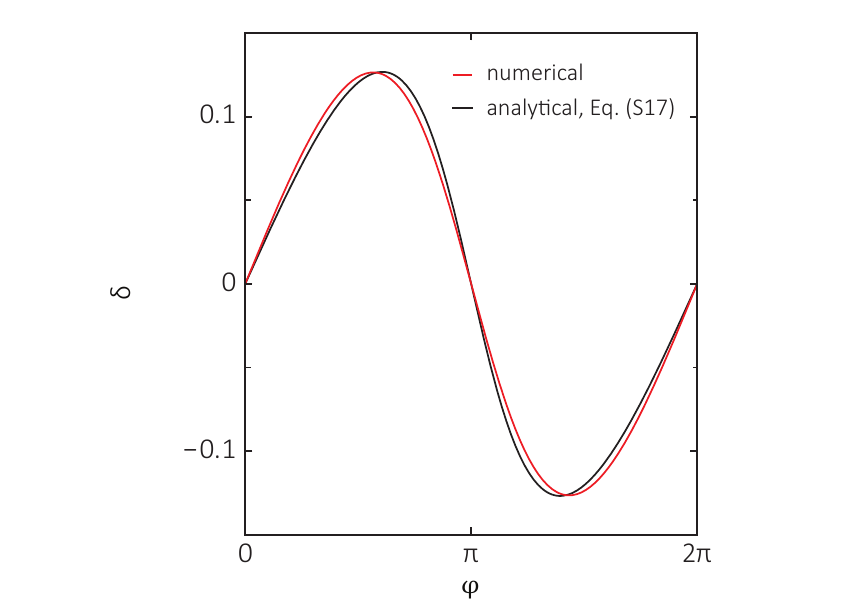}
\caption{\textbf{Equilibrium phase drop $\delta$ across the tunnel
junction.} The black line is given by Eq.~\eqref{eq:phase_drop_equilibrium_estimate}, the
red line by the numerical solution of Eq.~\eqref{eq:phase_drop_condition}. In
both cases, we use the same circuit parameters as in Fig.~3a of the main text:
$\Delta=122\,\mu$eV, $T=0.57$, $E_J=165\,\mu$eV.}
\end{figure}

Since  $\phi =\varphi - \delta$, (see Eq.~(\ref{eq:H_squid})), it is sufficient
to show that the equilibrium phase drop $\delta \equiv \langle GS| \hat{\delta}|
GS \rangle$ across the tunnel junction is small.
$\delta$ is given by the position where the ground
state Josephson energy of Eq.~\eqref{eq:H_squid} is minimal for $E_C=0$. From
this condition, after taking a derivative of the Josephson energy, we obtain the
following transcendental equation for $\delta$:
\begin{equation}\label{eq:phase_drop_condition}
E_J \sin(\delta) + \frac{\Delta\,T}{4}\,\frac{\sin(\delta-\varphi)}{\sqrt{1-T\sin^2[(\varphi-\delta)/2]}}=0\,.
\end{equation}
We note that the above expression defines a zero net current through the hybrid
SQUID with the two arms hosting the same supercurrent. For $E_J\gg\Delta T/4$, a
good approximate solution is given by
\begin{equation}\label{eq:phase_drop_equilibrium_estimate}
\delta \approx \frac{\Delta\,T}{4E_J}\,\,\frac{\sin(\varphi)}{\sqrt{1-T\sin^2(\varphi/2)}}\,.
\end{equation}
up to quadratic corrections in $(\Delta T/E_J)$. In Fig.~S9 we show that for the
parameters used in Fig.~3a, this approximate solution is very close to 
the exact, numerical one. Both exhibit a sinusoidal behavior with a
maximum $\delta\approx 0.12$ at $\varphi\approx \pi/2$. This confirms that the phase
drop across the weak link, $\phi=\varphi-\delta$, remains very close to the
applied phase $\varphi$ everywhere. In particular, $\phi$ is exactly
equal to $\varphi$ at $\varphi=n\pi$, where $n$ is integer.

\subsection{Andreev bound states in a proximitized Rashba nanowire in a parallel magnetic field}

In this Section, we introduce the model used to describe the behavior of ABS
as a function of the magnetic field $B$. We start from the standard
Bogoliubov-de Gennes (BdG) Hamiltonian of a Rashba quantum wire with
proximitized $s$-wave superconductivity and an external Zeeman field
\citeS{suppPhysRevLett.105.077001,suppPhysRevLett.105.177002}:
\begin{equation}\label{eq:Hbdg}
H_\textrm{BdG} = -\left(\frac{\pd_x^2}{2m} -
E_F\right)\,\tau_z\,-i\,\alpha\,\pd_x\,s_z\,\tau_z + E_Z\,s_x + \Delta\,e^{i\phi\,\theta(x)\,\tau_z}\,\tau_x + V\,\delta(x)\,\tau_z\,.
\end{equation}
Here, the two sets of Pauli matrices $\tau_{x,y,z}$ and $s_{x,y,z}$ act in the
Nambu and spin spaces, respectively;  $m=0.023 m_\textrm{e}$ is the effective
mass in InAs \citeS{supp10.1063/1.1368156}, $\alpha$ is the Rashba spin-orbit
coupling strength which defines $E_\textrm{SO}=m\alpha^2/2$. ${E_Z =
\tfrac{1}{2}g \mu_B B}$ is the Zeeman energy, $\Delta$ is the proximity
induced gap and $\theta$ is the Heaviside step function. The Fermi level $E_F$
is measured from the middle of the Zeeman gap in the normal state band
dispersion, see Fig.~S11. Note that starting with Eq.~\eqref{eq:Hbdg} we set
$\hbar=1$. The superconducting phase difference between the left lead ($x<0$) and the right
lead ($x>0$) is denoted by $\phi$. The last term of Eq.~(S18) models a
short-range scatterer at $x=0$, accounting for the finite channel transmission.

We seek bound state solutions of the the BdG equations,
\begin{equation}\label{eq:Bdg_eq_second_order}
H_\textrm{BdG}\,\Psi(x) = E\,\Psi(x)\,,
\end{equation}
at energies $|E|<\Delta$. We will consider in particular two opposite regimes:
(a) $E_F\gg E_\textrm{SO}, E_Z,\Delta$ and (b) $E_F=0$, see the two insets in
the corresponding panels of Fig.~S11. In order to find bound state solutions we
proceed as follows:
\begin{enumerate}
	\item We linearize the BdG equations for the homogeneous system ($V=0,
	\phi=0$) around $E=E_F$. In this way, we obtain two effective low-energy
	Hamiltonians, $H^{(a)}_\textrm{eff}$ and $H^{(b)}_\textrm{eff}$, which are
	linear in the spatial derivative. They can be written as:
\begin{subequations}\label{eq:Heff}
\begin{align}
H^{(a)}_\textrm{eff}&=-i v\, \pd_x\, \tau_z\,\sigma_z - v q_0\,\tau_z\,\rho_z + \frac{\Delta \,\alpha\,k_F}{v q_0}\,\,\tau_x \sigma_z + \frac{\Delta\,E_Z}{v q_0}\,\tau_y\,\rho_y\,,\\
H^{(b)}_\textrm{eff}&=-i \alpha\, \pd_x\, \tau_z\,\sigma_z + \Delta \tau_x + \tfrac{1}{2}\,E_Z\,\sigma_z\,(1-\rho_z)\,.
\end{align}
\end{subequations}
We now have three sets of Pauli matrices: $\tau_{x,y,z}$ (Nambu space),
$\rho_{x,y,z}$ [distinguishing the inner/outer propagating modes, and replacing
the spin matrices $s_{x,y,z}$ of Eq.~\eqref{eq:Hbdg}], and $\sigma_{x,y,z}$
(distinguishing left- and right-moving modes, and not to be confused with the
$\sigma$ matrices used in the previous Section). For regime (a), we have also
introduced the Fermi momentum $k_F=\sqrt{2mE_F}$, the Fermi velocity $v = k_F
/m$ and the energy difference  $vq_0=\sqrt{\alpha^2k_F^2 + E_Z^2}$ between the
two helical bands at the Fermi momentum. Note that, in the regime (b) where
$E_F=0$, the linearization requires $E_\textrm{SO}\gg \Delta, E_Z$, so it corresponds to
the limit of strong spin-orbit coupling.
\item Using Eq.~\eqref{eq:Hbdg}, we compute the transfer matrix $\mathcal{T}$ of
the junction in the normal state ($\Delta=0$), at energy $E=E_F$. The transfer
matrix gives a linear relation between the plane-wave coefficients of the
general solution on the left and right hand sides of the weak link. In computing
$\mathcal{T}$, we neglect all terms $\propto E_F^{-1}$ in regime (a). In regime
(b), the transfer matrix is computed for $E_Z=0$, since the effect of magnetic
field on scattering can be neglected to due the small dwell time in the short
junction. At $E_Z=0$, the transfer matrix depends on the single real parameter
$T$, the transmission probability of the junction. The latter is given by
$T=4k_F^2/(4k_F^2 + V^2)$ in regime (a), and $T=1/(1+V^2/\alpha^2)$ in regime
(b).
\item Using the transfer matrix $\mathcal{T}$ as the boundary condition at $x=0$
for the linearized BdG equations, we obtain the following bound state equation
for $E$:
	\begin{equation}\label{eq:transfer_matrix_equation}
	\det\,\left[1 - G(E)\,\tau_z\,\sigma_z\,\left(e^{-i\phi\,\tau_z/2}\,\mathcal{T}-1\right)\right]=0\,,
	\end{equation}
	where $G(E)$ is the integrated Green's function,
	\begin{equation}
	G(E) = v\,\int\frac{\de q}{2\pi i}\,e^{-iq\cdot0}\,[H_\textrm{eff}(q)-E]^{-1}\,,
	\end{equation}
and $H_\textrm{eff}(q)$ is the Fourier transform of either of the linearized
Hamiltonians of Eq.~\eqref{eq:Heff}. [In regime (b), $v$ must be replaced by
$\alpha$ in the expression for $G(E)$]. In deriving the bound state equation, we
have neglected the energy dependence of the transfer matrix, which is
appropriate in the short junction limit. In regime (b), this also requires that
the length of the junction is shorter than $\alpha/E_Z$, so that we can neglect
resonant effects associated with normal-state quasi-bound states in the Zeeman
gap, which would lead to a strong energy dependence of the transmission
\citeS{suppPhysRevB.93.174502}. Eq.~\eqref{eq:transfer_matrix_equation} is analogous
to the bound state equation for the ABS derived in
Ref.~\citeS{suppPhysRevLett.67.3836}, except that it is formulated in terms of the
transfer matrix of the weak link, rather than its scattering matrix. Unlike its
counterpart, Eq.~\eqref{eq:transfer_matrix_equation} incorporates the effect of
the magnetic field in the superconducting leads. It is thus appropriate to study
the effect of a magnetic field on the ABS in the limit of uniform penetration of
the field in the superconductor.
\item After performing the integral for $G(E)$, the roots of
Eq.~\eqref{eq:transfer_matrix_equation} can be determined numerically. For the
two regimes, this leads to the typical behavior of the ABS shown in Fig.~S11
against the experimental data. As mentioned in the main text and discussed
below, we find a better agreement with the experimental data for regime (a).
\end{enumerate}

From $G(E)$, we can also compute the proximity-induced gap of the continuous
spectrum $\Delta(B)$: $\Delta(B)$ is the minimum value of $E$ such that the
poles of $G(E)$ touch the real axis in the complex plane [of course, $\Delta(B)$
can also be found by minimizing the dispersion relation obtained by
diagonalizing Eq.~\eqref{eq:Heff} in momentum space]. In regime (a), the
relevant spectral gap is always at the finite momentum, so the behavior of $\Delta(B)$
depends on the strength of the spin-orbit coupling, as shown in Fig.~S10. Two
features are evident from the figure.

First, with increasing spin-orbit coupling, the linear behavior
$\Delta(0)-\Delta(B) \propto B$ changes to to a quadratic suppression
$\Delta(0)-\Delta(B) \propto B^2$ for small $B$.
This is due to the vanishing first-order matrix elements of the Zeeman
interaction, due to the removal of the spin degeneracy of finite-momentum states
by the spin-orbit interaction. Secondly, the proximity-induced gap $\Delta(B)$
never closes -- as long as the superconductivity in the aluminium shell is
present -- because spin-orbit interaction competes with the Zeeman effect and
prevents the complete spin polarization of the electrons.
These two facts explain the behavior of $\Delta(B)$ shown in Fig.~4b of the main
text.
In regime (b) with $E_F=0$, which is extensively discussed in the literature of
Majorana bound states, $\Delta(0)-\Delta(B)\propto B$ due to the Zeeman-induced
suppression of the gap for states at zero momentum (where spin-orbit is not
effective).

An in-depth theoretical study of Eq.~\eqref{eq:transfer_matrix_equation},
including a detailed analysis of its roots at finite magnetic fields and the
code used in the numerical solution, is in preparation. It will also be
interesting to extend the current model beyond the linearization to allow the
calculation of the spectrum at arbitrary values of $E_F$.

\begin{figure}[H]
\centering
\includegraphics[width=0.45\textwidth]{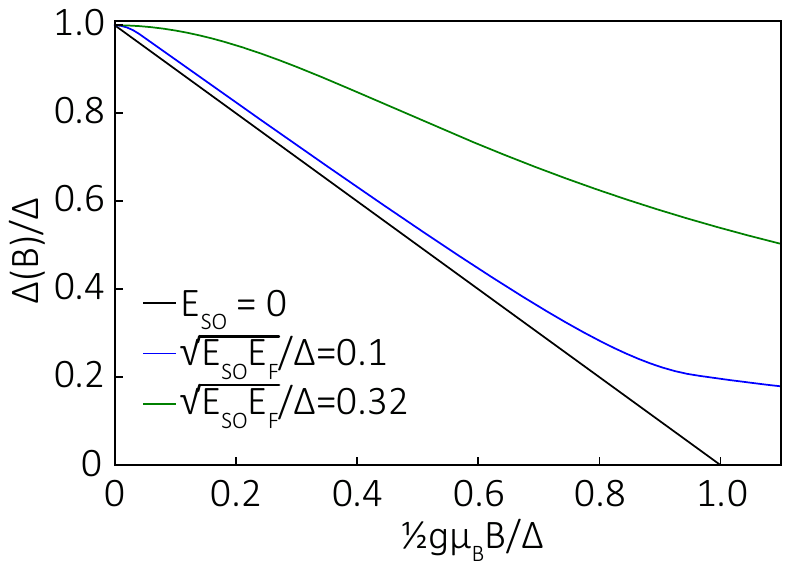}
\caption{\textbf{The effect of the spin-orbit interaction and Zeeman field on
the induced superconducting gap.} The lack of spin-orbit interaction leads to 
a linear decrease of $\Delta(B)$ (black line), which becomes parabolic in the
limit of $\sqrt{E_\textrm{SO}E_F} \gg E_Z=\frac{1}{2}g\mu_B B$ (blue and green
lines). The green line corresponds to the best fit to the experimental data
shown in the main text.}
\end{figure}

\subsection{Orbital field}

Because a quadratic suppression of $\Delta(B)$ and the ABS energies may also be
due to the orbital effect of the magnetic field, without invoking spin-orbit
interaction, it is important to compare the data with this scenario. In a simple
model which includes orbital and Zeeman effect, the field-dependence of the
Andreev bound states may be written down as follows:
\begin{equation}\label{eq:orbital_abs}
E^\textrm{(orb)}_{\textrm{ABS},\pm}(\phi, B) = \Delta (1-B^2/B_*^2)\,\sqrt{1-T \sin^2(\phi/2)}\pm (1/2) g \mu_B B\,.
\end{equation}
Here, $B_* \sim \Phi_0/A$ is the magnetic field scale which governs the
suppression of the proximity-induced gap due to the orbital field, $A$ is the
cross-section of the nanowire and $\Phi_0=h/2e$. In writing
Eq.~\eqref{eq:orbital_abs}, we have neglected the effect of the orbital field on
the scattering at the junction. This should be a good approximation as long as
the junction is modeled by a $\delta(x)$ potential with no dependence on the
radial coordinate of the nanowire. Thus, essentially, the phase dependent part
of the Andreev bound state energies can be obtained by replacing $\Delta$ with
$\Delta (1-B^2/B_*^2)$ in Eq.~(1) of the main text. In the absence of spin-orbit
coupling, the Zeeman term enters additively in Eq.~\eqref{eq:orbital_abs}.

Using Eq.~\eqref{eq:orbital_abs}, we can perform a fit to the experimental data
to determine the optimal value $B_*=400 \pm 2$ mT.
Note that the fit is insensitive to the value of $g$, since $g$ drops out from
the sum $E^\textrm{(orb)}_{\textrm{ABS},+} + E^\textrm{(orb)}_{\textrm{ABS},-}$.
However, Eq.~\eqref{eq:orbital_abs} predicts the occurrence of a fermion
parity-switch at a field $B_\textrm{sw} < B_*$ given by the condition
$E^\textrm{(orb)}_{\textrm{ABS},-}(\phi, B_\textrm{sw})=0$.
From this condition, and assuming the knowlede of both $B_\textrm{sw}$ and
$B_*$, the $g$-factor can then be deduced by inverting
Eq.~\eqref{eq:orbital_abs} at $\phi=\pi$,
\begin{equation}
g = \frac{\Delta\sqrt{1-T}}{\mu_B\,B_\textrm{sw}}\,\left(1-B_\textrm{sw}^2/B^2_*\right)
\end{equation}
The occurrence of this fermion-parity switch must be accompanied by a drastic
disappearance of the ABS transition \citeS{suppPhysRevB.77.184506}.
In the experiment, such disappearance can be excluded up to at least $300$ mT.
Therefore, by requiring that $B_\textrm{sw}>300$ mT and using the values quoted
in the main text for all other parameters, we obtain an upper bound of $g$,
\begin{equation}
\abs{g}<5.08
\end{equation}

In Fig.~S11c we plot the energy spectrum resulting from Eq.~(S23), which
includes only the orbital and Zeeman effects.
The black line in Fig.~S11c represents the edge of the continuous spectrum for
states with spin down, $\Delta(B)=\Delta (1- B^2/B_*^2) - \tfrac{1}{2} g \mu_B
B$. In Fig. S11c, we choose $g=5$, close to the upper bound of
Eq.~(S25). The inclusion of a weak spin-orbit coupling in the model would
not affect the curvature of $\Delta(B)$ and
$E^{\textrm{(orb)},\pm}_{\textrm{ABS}}(B)$ at small fields $g\mu_B B\ll \Delta$
(see the blue curve in Fig.~S10): the curvature would still be entirely dictated by the orbital effect.
As mentioned in the main text, the Andreev level and the continuum cross at a
value of the field $B_\textrm{cross}$ such that
$E_\textrm{ABS}^\textrm{(orb),+}(B_\textrm{cross})=\Delta(B_\textrm{cross})$.
For $B_*=400\,$mT and $g=5$, the crossing happens at $B_\textrm{cross}\approx
150\,$mT, see Fig.~S11c.
However, the inclusion of a weak spin-orbit coupling prevents the level
crossing, causing the Andreev level to bend below the edge of the continuum.
As a consequence, the transition energy $E_\textrm{tot}(B)$ decreases
sharply at $B>B_\textrm{cross}$, in contrast with its behavior in the absence of
spin-orbit coupling (compare the dashed and dotted lines in Fig.~S11c).
The behavior of $E_\textrm{tot}(B)$ in the presence of weak spin-orbit coupling
clearly disagrees with the experimental data in the field range $150\,$mT$<B<
300\,$mT.

The considerations above motivate the approximation used in the main text, where
we attribute the quadratic suppression of $E_\textrm{tot}(B)$ to the joint
effect of spin-orbit and Zeeman couplings; the orbital effect does not play a
dominant role in the observed dispersion.

\subsection{Fits to the data}

We have presented three different scenarios that can be used to interpret
the magnetic field dependence of the ABS transition energies. We have fitted all
three models to the entire data set available, consisting of a flux bias sweep
of the ABS spectra at six different magnetic fields ($B=50, 75, 100, 150, 200$
and $300$ mT). For each flux bias at which it was visible, we have extracted the
position of the ABS transition. For each value of $B$ we attributed to all the
data points an error bar corresponding to the half-width at half-maximum of the
ABS peak at $\varphi=\pi$, neglecting for simplicity the flux variation of the
width. The total dataset consisted of more than 300 datapoints. We then
performed a least-square fit to the ABS transition energies predicted by the
three different models. The results are illustrated in Fig.~S11.

\newpage

\begin{figure}[H]
\centering
\includegraphics[width=0.85\textwidth]{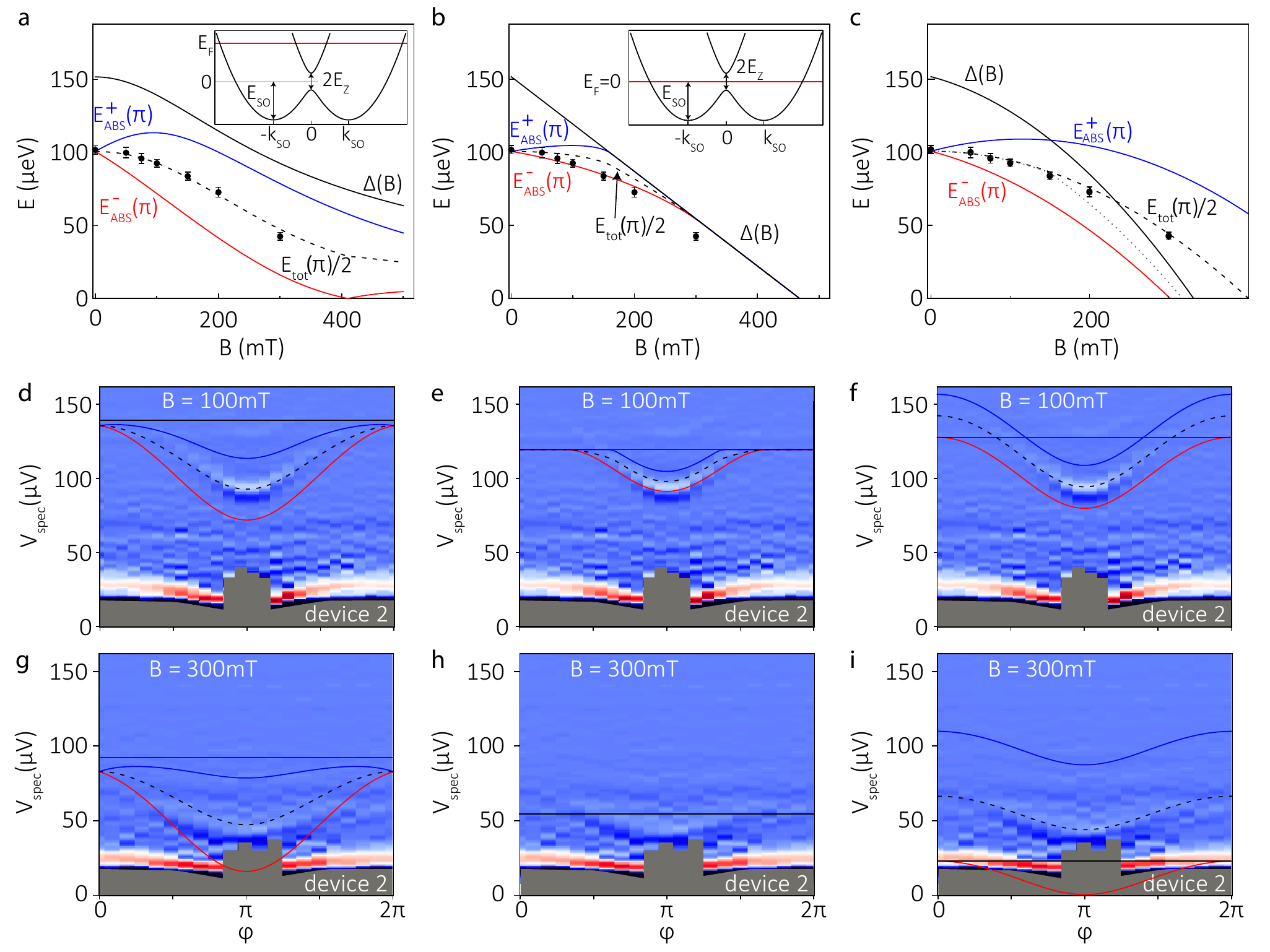}
\caption{\textbf{The magnetic field dependence of ABS in high and low Fermi
level regimes and for orbital magnetic field.} The top row shows the evolution
of the spin-split Andreev levels $E_\textrm{ABS}^\pm(B)$ (blue and red lines),
the transition energy
$E_\textrm{tot}(B)=E_\textrm{ABS}^+(B)+E_\textrm{ABS}^-(B)$ (dashed line) at
$\varphi=\pi$ and the proxitimity-induced gap $\Delta(B)$ versus the magnetic
field $B$. The theoretical curves for $E_\textrm{tot}$ are compared against the
experimental data (black dots). The three panels correspond to the three
different theoretical models described in the text: high Fermi level (a), low
Fermi level (b), and a model without spin-orbit coupling but only orbital and
Zeeman effects of the field (c). For the latter, the dotted line in panel (c)
depicts the qualitative behavior of $E_\textrm{tot}$ if a weak-spin orbit
coupling is included in the model.
The middle and the bottom row show the resulting dispersion of the Andreev
levels as a function of $\varphi$ for the three different theoretical models,
displayed on top of the measured spectrum at $B=100\,$mT and $B=300\,$mT,
respectively. In each row, all three columns feature the same experimental
dataset. The global fit parameters for the left column are $g=14.7\pm0.6$ and
$\sqrt{E_\textrm{SO}E_F}/\Delta = 0.32\pm 0.02$. The middle column is evaluated
with a single fit parameter $g=11.2\pm0.1$. Note the lack of dispersion in panel
(h), due to the merging of the Andreev bound states with the continuum, which
causes all the lines to fall on top of each other. In the right column we use
the best-fit value $B_*=400\pm2\,$mT and $g=5$, the latter imposed by the lower
bound on the parity switching field $B_\textrm{sw} > 300\,$mT, where
$E_\textrm{ABS}^-(\pi)=0$.}
\end{figure}

\bibliographystyleS{apsrev4-1}
\bibliographyS{references_ABS_nourl_suppl}

\end{document}